\documentclass[showpacs,10pt,aps,prd,reprint,onecolumn,notitlepage,nofootinbib,superscriptaddress]{revtex4-1}

\usepackage{hyperref}
\usepackage{amsmath}
\usepackage{mathtools}
\usepackage{bm}
\usepackage{amssymb}
\usepackage{graphicx}
\usepackage[]{subfigure}
\usepackage{filecontents}
\usepackage[dvipsnames]{xcolor}

\hypersetup{
    bookmarks=true,         % show bookmarks bar?
    pdftoolbar=true,        % show Acrobats toolbar?
    pdfmenubar=true,        % show Acrobats menu?
    pdffitwindow=true,     %window fit to page when opened
    pdfstartview={FitH},     %fits the width of the page to the window
    pdftitle={My title},     %title
    pdfauthor={author},      %author
    pdfsubject={Subject},    %subject of the document
    pdfcreator={Creator},    %creator of the document
    pdfproducer={Producer},  %producer of the document
    pdfkeywords={keyword1} %{key2} {key3}, % list of keywords
    pdfnewwindow=true,       %links in new window
    colorlinks=true ,        %false: boxed links; true: colored links
    linkcolor=blue  ,        %color of internal links (change box color with linkbordercolor)
    citecolor=blue,      %color of links to bibliography
    filecolor=green,       %color of file links
    urlcolor=blue            %color of external links
}

\newcommand{\arcsinh}{\mathrm{arcsinh}}
\newcommand{\arccosh}{\mathrm{arccosh}}
\newcommand{\rO}{\mathrm{\Omega}}
\newcommand{\ed}{\mathrm{d}}

\begin{document}

\title{Gravitational Rutherford scattering of electrically charged particles from a charged Weyl black hole}

\author{Mohsen Fathi}
\email{mohsen.fathi@postgrado.uv.cl}
\affiliation{Instituto de F\'isica y Astronom\'ia, Universidad de Valpara\'iso, Avenida Gran Breta\~na 1111, Valpara\'iso, Chile}

\author{Marco Olivares}
\email{marco.olivaresr@mail.udp.cl}
\affiliation{ Facultad de  Ingenier\'{i}a y Ciencias, Universidad Diego Portales, Avenida Ej\'{e}rcito            Libertador 441, Casilla 298-V, Santiago, Chile}

\author{J.R. Villanueva}
\email{jose.villanueva@uv.cl}
\affiliation{Instituto de F\'isica y Astronom\'ia, Universidad de Valpara\'iso, Avenida Gran Breta\~na 1111, Valpara\'iso, Chile}
\begin{abstract}
Considering electrically charged test particles, we continue our study of the exterior dynamics of a charged Weyl black hole which has been previously investigated regarding the motion of mass-less and (neutral) massive particles. In this paper, the deflecting trajectories of charged particles are designated as being gravitationally Rutherford-scattered and detailed discussions of angular and radial particle motions are presented.

\end{abstract}

\maketitle

%%%%%%%%%%%%%%%%%%%%%%%%%%%%%%%%%%%%%%%%%%%%%%%%%
\section{Introduction}

{The scattering of charged particles in electric fields is indeed one of the most re-known phenomena in physics and has had numerous applications in small and large scale observations. Regarding the former, and without loss of generality, the famous Rutherford scattering experiment that led to the discovery of the atomic nucleus, is described in terms of elastic deflecting trajectories of charged particles from a heavy charged central mass. Such particle trajectories, beside being well-known in small atomic scales, have been also investigated widely in black hole spacetimes. In fact, the study of motion of test particles in the gravitational field of black holes, dates back to the early days of general relativity and ever since, it has found its way in classic textbooks (see for example Refs.~\cite{Misner:1973,Futterman:1988,Chandrasekhar:579245} and the reviews in Refs.~\cite{Poisson:2011,Blanchet:2011}). The interest in performing such studies, beside their applicability in testing general relativity and modified theories of gravity, stems mostly in the opportunity that they provide to correctly analyze the dynamics of extremely warped regions around black holes. In these regions, based on the effective gravitational potential that affects the particles, they can lie on different types of orbits, among which, and in particular, the deflecting trajectories relate tightly to the scattering phenomena. It is well-known that the charge parameter of charged black hole spacetimes (like Reissner-Nordstr\"{o}m and Kerr-Newman), contributes in the gravitational potential of the black hole and therefore, can affect the motion of neutral particles. In the case of charged test particles moving around such black holes, the additional electromagnetic potential changes the nature of deflecting trajectories to a special form of the Rutherford scattering. The importance of this kind of motion is such that it has received a large number of performed studies in analyzing, numerically and analytically, the respected equations of motion and the scattering cross-sections. These studies have been done in the contexts of general relativity and alternative gravity (see for example Refs.~\cite{Bicak:1989,Karas:1990,Aliev:2002,Pugliese:2011,Olivares:2011,Fathi:2013,Hackmann:2013,Lim:2015,Garcia:2015,Pugliese:2017,Das:2017,Iftikhar:2018,Vrba:2020,Khan:2020,Yi:2020,Abdujabbarov:2020,Javlon:2020,Anacleto:2020} and Refs.~\cite{Villanueva:2015,Sarkar:2018,Zhao:2018,Gonzales:2018,Shaymatov:2020,Narzilloev:2020}). Moreover, regarding the chaotic nature of particle scattering, discussed in Ref.~\cite{Stuchlik:2016}, realistic astrophysical situations can be found that also demonstrate the creation of ultra-high energy particles \cite{Stuchlik:2020,Stuchlik:2020b}.

Although black holes with net electric charge are still remained as purely theoretical objects, however, studying them can pave the way in understanding physical phenomena like radiation reaction of particles \cite{Gal:1982,Tursunov:2018} and black hole evaporation \cite{Chen:2019}. Hence, the interest in investigating particle motion around charged black holes becomes more justified and, as well as in general relativity, it has found its way into alternative theories of gravity.}

Along the same effort, we investigate the motion and the scattering of charged test particles, as they travel in a particular charged black hole spacetime, which has been obtained as a non-vacuum solution to the Weyl conformal theory of gravity. In this discussion, we focus on studying the possible radial and angular motions of charged test particles that approach the aforementioned charged Weyl black hole (CWBH). Our method of study is based on the standard Hamilton-Jacobi formulation of particle motion. In particular, we concern with the deflecting trajectories which are strongly related to the gravitational version of Rutherford scattering of charged particles. The equations of motion are those corresponding to hyperbolic orbits and it is found that, finding analytical solutions to these equations, demands specific mathematical tricks. Aside form this, we also calculate the Rutherford scattering for particles on radial trajectories and the evolution of temporal parameters are derived. Additionally, assuming the congruence deviation of a bundle of infalling world-lines, we discuss the internal interactions between the particles and point out their effects on the kinematical congruence expansion. 

The paper is organized as follows: In Sec.~\ref{sec:Weyl}, we give a review on Weyl conformal gravity and bring in the black hole solution that we are intended to investigate. In Sec.~\ref{MCP}, the basic equations governing the motion of charged particles in the spacetime generated by the CWBH, are given by means of the Hamilton-Jacobi formalism. In particular, the angular motion is analyzed in detail in Sec.~\ref{sec:angular}, which is followed by that for the radial motion in Sec.~\ref{sec:radial}. In these two sections, the gravitational Rutherford scattering is formulated analytically and the trajectories are {plotted} respectively. In Sec.~\ref{sec:deviation}, a bundle of particle world-lines is considered and we use its deviation as a tool to discuss the congruence's internal acceleration and expansion. We close our discussion in Sec.~\ref{sec:conclusion}.

%%%%%%%%%%%%%%%%%%%%%%%%%%%%%%%%%%%%%%%%%
\section{The Weyl conformal theory of gravity and its static black hole solutions}\label{sec:Weyl}

{
To obtain the Weyl conformal theory of gravity, the common Einstein-Hilbert action of general relativity
\begin{equation}\label{eq:IEH}
    I_{EH} = \mathcal{K}_{EH}\int \ed^4 x\sqrt{-g}~R,
\end{equation}
is replaced by
\begin{equation}\label{eq:IWeyl}
    I_W=-\mathcal{K}_W\int{\ed^4x\sqrt{-g}\,\,C_{\alpha\beta\mu\nu}C^{\alpha\beta\mu\nu}}.
 \end{equation}
In the above actions, $\mathcal{K}_{EH}$ and $\mathcal{K}_W$ are appropriate coupling constants, $g=\mathrm{det}(g_{\mu\nu})$, $R = g^{\nu\mu}  {R^{\alpha}}_{\mu\alpha\nu} = {R^\nu}_\nu$ is the Ricci curvature scalar, and the Weyl tensor
\begin{equation}\label{eq:C}
C_{\alpha\beta\mu\nu} = R_{\alpha\beta\mu\nu}-\frac{1}{2}\left(g_{\alpha\mu}
R_{\beta\nu}-g_{\alpha\nu}R_{\beta\mu}-g_{\beta\mu}R_{\alpha\nu}+g_{\beta\nu}R_{\alpha\mu}\right)\\
+\frac{R}{6}\left(g_{\alpha\mu}g_{\beta\nu}-g_{\alpha\nu}g_{\beta\mu}\right),
\end{equation}
is invariant under the conformal transformation $g_{\mu\nu}(x) = e^{2 \ell(x)} g_{\mu\nu}(x)$, where $2 \ell(x)$ is the local spacetime stretching. Accordingly, the action \eqref{eq:IWeyl} can be recast as
\begin{equation}\label{eq:IWeyl-2}
    I_W=-\mathcal{K}_W\int
\ed^4x
\sqrt{-g}\,\left(R^{\alpha\beta\mu\nu}R_{\alpha\beta\mu\nu}-2R^{\mu\nu}R_{\mu\nu}+\frac{1}{3}R^2\right),
\end{equation}
in which, the Gauss-Bonnet term $\sqrt{-g}\,(R^{\alpha\beta\mu\nu}R_{\alpha\beta\mu\nu}-4R^{\mu\nu}R_{\mu\nu}+R^2)$
is a total divergence and hence, the action is simplified as \cite{Mannheim:1989,Kazanas:1991}
\begin{equation}\label{eq:IWeyl-3}
    I_{W}=-2\mathcal{K}_W\int{\ed^4x}\sqrt{-g}\,\,\left(R^{\mu\nu}R_{\mu\nu}-\frac{1}{3}R^2\right).
\end{equation}
Applying $\frac{\delta{I_W}}{\delta{g_{\mu\nu}}} = 0$, the equations of motion are derived as
\begin{eqnarray}\label{eq:Bach}
    W_{\mu\nu} &=& \mathfrak{P}_{\alpha\beta}\,{{{C_{\mu}}^{\alpha}}_{\nu}}^{\beta}+\nabla^\alpha \nabla_\alpha \mathfrak{P}_{\mu\nu}-\nabla ^{\alpha}\nabla _{\mu}\mathfrak{P}_{\nu\alpha}\nonumber\\
    &=& \frac{1}{4\mathcal{K}_W} T_{\mu\nu},
\end{eqnarray}
in which
\begin{equation}
    \mathfrak{P}_{\mu\nu}=\frac{1}{2}\left(R_{\mu\nu}-\frac{1}{4} g_{\mu\nu} R\right),
\end{equation}
is the  Schouten tensor. The field equation \eqref{eq:Bach} is known as the Bach equation. Based on the fact that Weyl gravity is a theory of fourth order in the metric, the Bach equation therefore contains derivatives up to the fourth order. Such field equations may lead to more convergent and renormalizable theories of gravity, although, the consistency of their quantization is still under debate, mostly because the field equations are conceived as fluctuations on a fixed background \cite{Mannheim:2007}. Additionally, the fourth-order theories of gravity usually give rise to the presence of ghost fields, although some solutions to this issue have been reported in Ref.~\cite{Mannheim:2006}. It is of worth mentioning that the standard model of particle physics can be endowed with conformal symmetry, if an appropriate gravitational term is added to its action. This gravitational coupling helps fixing a conformal gauge in accordance with a reference mass. This way, the action can generate particle mass, without encountering any symmetry breaking (see Ref.~\cite{Pawlowski:1994} for more details).}

{In 1989, Mannheim and Kazanas obtained a static spherically symmetric solution to the vacuum Bach equation ($T_{\mu\nu}=0$), which was given by the metric\footnote{In this paper, we work in the geometric unit system, by considering $G=c=1$. Accordingly, the dimensions of mass, electric charge and length are given in meters.}
\begin{equation}
	{\rm d}s^{2}=-B(r)\, {\rm d}t^{2}+\frac{{\rm d}r^{2}}{B(r)}+r^{2}({\rm d}\theta^{2}+\sin^{2}\theta\,
	{\rm d}\phi^{2}), \label{metr}
\end{equation} 
with the lapse function \cite{Mannheim:1989}
\begin{equation}\label{eq:originalWeylB(r)}
 B(r) = 1-\frac{\zeta\left(2-3\zeta\rho\right)}{r}  - 3\zeta \rho + \rho r - \sigma r^2. 
\end{equation}
The first two terms are in common with general relativistic vacuum solutions, in the sense that by letting $\zeta = M_0$ and $\rho=\sigma=0$, the Schwarzschild solution is obtained for a spherically symmetric source of mass $M_0$. The last two terms, however, are peculiar to the fourth order Bach equation. Essentially, this solution was intended to explain the flat galactic rotation curves \cite{Rubin1980}, by introducing an extra gravitational potential in the spacetime, and therefore, tries to recover one of the cosmological evidences that support the dark matter scenario. For a typical galaxy of about 10 kpc in length, this potential grows with distance for $r > 10$ kpc, is approximately constant for $r\sim 10$ kpc, and becomes Newtonian for $r < 10$ kpc (for this latter condition, see Refs.~\cite{Yoon:2013,Mannheim:2013} for more discussions). Accordingly, the estimated value of $\rho$, for which, the $r$ and $1/r$ related terms are comparable, is $\rho\approx 10^{-26} \mathrm{m}^{-1}$. This way, the solution encounters the dimensionless quantity $\zeta\rho\approx 10^{-12}$ and predicts the flat rotational velocity of $10^2 \mathrm{km}\mathrm{s}^{-1}$ \cite{Mannheim:1989}. Furthermore, the last term of the solution \eqref{eq:originalWeylB(r)}, accounts for supporting the accelerated expansion of the universe, and therefore is related to the dark energy scenario \cite{Mannheim:2005}. So, as it is observed, the theory generates a substantially different gravitational potential to general relativity. Moreover, in the case that the energy-momentum tensor is associated with the electrostatic vector potential \begin{equation}\label{eq:vectorPotential}
    A_\alpha = \left(
    \frac{q_0}{r},0,0,0
    \right),
\end{equation}
corresponding to a spherically symmetric massive source of electric charge $q_0$, the same line element as in Eq.~\eqref{metr} has been applied by Mannheim and Kazanas, to obtain the Reissner-Nordstr\"{o}m form of solution to Weyl gravity, with the lapse function \cite{Mannheim1991}
\begin{equation}\label{eq:Mannheim}
    B(r) = \mathfrak{w} + \frac{\mathfrak{u}}{r}+\mathfrak{v} r - \mathfrak{k} r^2,
\end{equation}
for which, the parameters can be rewritten as
\begin{subequations}\label{eq:wuv}
\begin{align}
 &   \mathfrak{w} = 1-3\zeta\rho,\\
 &   \mathfrak{v} = \rho,\\
 &   \mathfrak{u} = -\zeta\left(2-3\zeta\rho\right)-\frac{q_0^2}{8\rho \mathcal{K}_W},
\end{align}
\end{subequations}
in terms of the coefficients presented in the vacuum solution \eqref{eq:originalWeylB(r)}. In Ref.~\cite{Payandeh:2012mj}, the vector potential \eqref{eq:vectorPotential} has been applied, once again, to the Bach equation \eqref{eq:Bach}, however, for the reference lapse function{{
\begin{equation}\label{eq:F_metric_0}
    B(r) = 1+\frac{1}{3}\left(c_2 r + c_1 r^2\right),
\end{equation}
in which, the two coefficients $c_1$ and $c_2$ replace, respectively, the dark matter and dark energy terms, as in the lapse function \eqref{eq:originalWeylB(r)}. These coefficients can be determined by describing the solutions to the Bach equation, as perturbations on the Minkowski spacetime (appearing in the last two terms of Eq.~\eqref{eq:F_metric_0}). Accordingly, we can decompose the spacetime metric as $g_{\mu\nu} = h_{\mu\nu}+\eta_{\mu\nu}$, where $h_{\mu\nu}$ and $\eta_{\mu\nu}$ are respectively, the perturbation and the Minkowski metrics. Therefore, for a spherically symmetric source of mass $\tilde{m}$, electric charge $\tilde{q}$ and radius $\tilde{r}$, the Poisson equation for the perturbation field, i.e. $\nabla^2 h_{\mu\nu} = 8\pi T_{\mu\nu}$, yields
\begin{equation}\label{eq:Poisson_1}
    \nabla^2 h_{00} = 8\pi T_{00} = 8\pi\left( \frac{\tilde{m}}{\frac{4}{3}\pi \tilde{r}^3}+\frac{1}{8\pi}\frac{\tilde{q}}{r^4}\right), 
\end{equation}
as its 00 component. Here, $T_{00}$ corresponds to the scalar potentials produced by the mass and charge of the source. Applying Eq.~\eqref{eq:F_metric_0} in Eq.~\eqref{eq:Poisson_1}, we get  \cite{Payandeh:2012mj}
{\begin{equation}\label{eq:c1c2}
    c_2 = -\left(\frac{9\tilde{m}\,r}{\tilde{r}^3}+\frac{3}{2}\frac{\tilde{q}^2}{r^3}+3c_1 r\right),
\end{equation}}
that provides}}
\begin{equation}\label{lapse}
	B(r)=1-\frac{r^{2}}{\lambda^{2}}
	-\frac{Q^2}{4 r^2},
\end{equation}
in which
\begin{eqnarray}
&&\frac{1}{\lambda^2}=\frac{3 \tilde{m}}{\tilde{r}^{3}}   +\frac{2 \tilde\varepsilon}{3},\label{par1}\\
&&Q=\sqrt{2}\, \tilde{q}.\label{par2}
\end{eqnarray}
{Because we have excluded $c_2$ by defining it in terms of $c_1$, the lapse function \eqref{lapse} does not contain the dark matter related term, and $\tilde\varepsilon\equiv c_1$ in Eq.~\eqref{par1}, adds a cosmological term to the spacetime. Regarding this, the lapse function \eqref{lapse} describes the exterior geometry of a CWBH with a cosmological component.} This spacetime has been recently investigated regarding the motion of mass-less and neutral massive particles \cite{Fathi:2020a,Fathi:2020d,Fathi:2020b}. When $\lambda>Q$, this spacetime admits an event and a cosmological horizon, which are obtained by solving $B(r) = 0$, and are given respectively by \cite{Fathi:2020b}
\begin{eqnarray}
&& r_+= \lambda \sin\left( {1\over 2} \arcsin\left(\frac{Q}{\lambda} \right)  \right),\label{w.6}\\
&& r_{++}=\lambda \cos\left( {1\over 2} \arcsin\left(\frac{Q}{\lambda}\right)\right).\label{w.7}
\end{eqnarray}
These horizons unify at the distance $\mathrm{r}_{\mathrm{ex}}=r_+=r_{++}=\lambda/\sqrt{2}$, in the case of an extremal CWBH, when $\lambda=Q$. Accordingly, a naked singularity is encountered for $\lambda<Q$. {Note that, although the solution is singular in its character, we however, are concerned with regions outside the even horizon (and therefore, outside the massive source), so that we do not encounter the singularity at the origin.} 

{Furthermore, Letting $\tilde{r}$ to be a variable radial distance, $r$, $3\tilde{m}\rightarrow M_0$, $2\tilde{\varepsilon}\rightarrow\pm \Lambda$, and $Q\rightarrow 2 i Q_0$, we get to the Reissner–Nordstr\"{o}m–(Anti-)de Sitter black hole of mass $M_0$, charge $Q_0$ and cosmological constant $\Lambda$. However, as it is noticed, this transition requires an imaginary transformation. We begin studying the motion of charged test particles around the CWBH, from the next section, by employing the Hamilton-Jacobi formalism.}}

%%%%%%%%%%%%%%%%%%%%%%%%%%%%%%%%%%%%%%%
\section{Motion of charged particles around the CWBH}\label{MCP}

The Hamilton-Jacobi method of describing the motion of particles of mass $m$ and charge $q$ in an electromagnetic field, is based on the superhamiltonian \cite{Misner:1973}
\begin{equation}\label{eq:superhamiltonian}
    \mathcal{H} = \frac{1}{2} g^{\mu\nu} p_\mu p_\nu,
\end{equation}
in which the 4-momentum $\bm{p}$ satisfies $p_\mu p^\mu = -m^2$ and is defined as
\begin{equation}\label{eq:4-momentum}
    p_\mu = g_{\mu\nu} \frac{\ed x^\nu}{\ed\tau} = \left(
    \pi_\mu + q A_\mu
    \right),
    \end{equation}
in terms of the affine parameter $\tau$, the vector potential $\bm{A}$ and the generalized momentum $\bm{\pi}$, which is given according to the canonical Hamilton equation
\begin{equation}\label{eq:HamiltonEq_1}
    \frac{\ed\pi_\mu}{\ed\tau} = -\frac{\partial\mathcal{H}}{\partial x^\mu}.
\end{equation}
Recasting $\mathcal{H}$ in terms of the {\it characteristic Hamilton function} (i.e. the Jacobi action) 
\begin{equation}\label{eq:actionS}
   \mathcal{H} = - \frac{\partial S}{\partial\tau},
\end{equation}
we have $\pi_\mu = \partial S/\partial x^\mu$ and the Hamilton-Jacobi equation of the wave crests can be written as
\begin{equation} 
{1\over 2}g^{\mu \nu}\left(\frac{\partial S}{\partial
	x^{\mu}}+q A_{\mu}\right)\left(\frac{\partial S}{\partial
	x^{\nu}}+q A_{\nu}\right)+\frac{\partial S}{\partial\tau}=0.
\label{mcp1}
\end{equation}
The generalized momentum $\bm{\pi}$ is indeed responsible for the possible constants of motion. For stationary spherically symmetric spacetimes, such as that in Eq.~\eqref{metr}, these constants are
\begin{subequations}\label{eq:pi_t,pi_phi}
\begin{align}
    &\pi_t \doteq - E = g_{tt} \frac{\ed t}{\ed\tau} - q A_t,\\
    &\pi_\phi \doteq L = g_{\phi\phi} \frac{\ed\phi}{\ed\tau} - q A_\phi.
\end{align}
\end{subequations} \\
The constant $L$ corresponds to the particles' angular momentum and for motion in asymptotically flat spacetimes, $E$ is their energy. Now, confining the motion to the equatorial plane ($\theta = \pi/2$) and taking into account the only non-zero term of the vector potential of a CWBH (i.e. $A_t = \frac{\tilde{q}}{r}=\frac{Q}{\sqrt{2} r}$), we can specify Eq. (\ref{mcp1}) as
\begin{equation} 
\frac{-1}{B(r)}\left(\frac{\partial S}{\partial t}+\frac{q Q
}{\sqrt{2} r}\right)^{2}+B(r)\left(\frac{\partial S}{\partial
	r}\right)^{2}+\frac{1}{r^{2}}\left(\frac{\partial
	S}{\partial \phi}\right)^{2}+2\frac{\partial S}{\partial\tau}=0.
\label{mpc2}
\end{equation}
Based on the method of separation of variables of the Jacobi action, Eq.~\eqref{mpc2} can be solved by defining \cite{Carter:1968}
\begin{equation}\label{eq:action_carter}
    S=-E t+S_{0}(r)+L \phi+{1\over 2}m^{2}\tau,
\end{equation}
for which, interpolation in Eq.~\eqref{mpc2} results in
\begin{equation} 
S_{0}(r)= \pm \int \frac{\ed r}{B(r)} \sqrt{(E-V_-)(E-V_+)},
\label{mpc3}
\end{equation} 
where the radial potentials
are given by
\begin{subequations}\label{mpc3.1}
\begin{align}
    &V_{\pm}(r)=V_q(r)\pm\sqrt{B(r) \left( m^{2}+
	\frac{L^{2}}{r^{2}}\right)},\\
	 &V_q(r)\doteq\frac{q Q}{\sqrt{2} r}.
\end{align}
\end{subequations}
Note that, both of the $\pm$ branches of $V_\pm(r)$ converge to
the value $E_+=\frac{q Q
}{\sqrt{2} r_+}$ at $r=r_+$, which can be either positive or negative, depending
on the sign of the electric charges, $q$ and $Q$. 
Here we adopt the condition $q Q>0$, so that the $V_-$ branch is always negative (in the causal region $r_+<r<r_{++}$), and we
can consider the positive branch as the effective potential, i.e.
$V_{eff} \doteq V_{+}\equiv V$. 
{Furthermore, applying Eqs.~\eqref{eq:action_carter} and \eqref{mpc3}, it is possible to obtain the following three velocities:}
\begin{equation}\label{veltau}
u(r)\equiv \frac{\ed r}{\ed\tau}=\pm \sqrt{(E-V_-) (E-V)},
\end{equation}
\begin{equation} 
v_{t}(r)\equiv\frac{\ed r}{\ed t}=\pm  \frac{B(r) u(r)}{E-V_{q}\left(r\right)},
\label{mpc5}
\end{equation}
\begin{equation}\label{velphi}
v_{\phi}(r)=\frac{\ed r}{\ed\phi}=\pm \frac{r^{2} u(r)}{L}.
\end{equation}
The zeros of the above velocities do correspond to the so-called turning points, $r_t$, which are specified by the condition $V(r_t)=E_t$. Additionally, these equations lead to the quadratures that determine the evolution of the trajectories. This is dealt with in the forthcoming sections and the corresponding analytical solutions are obtained. In the next section, we begin with the angular motion of the test particles.

%%%%%%%%%%%%%%%%%%%%%%%%%%%%%%%%%%%%%%%%%%%%%%%%%%%
\section{Angular Motion}\label{sec:angular}

\begin{figure}[t]
	\begin{center}
		\includegraphics[width=8.5cm]{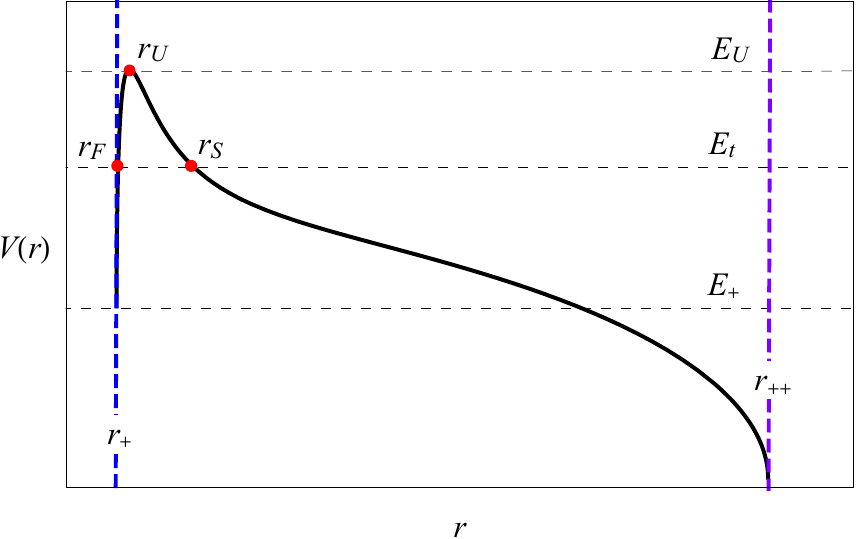}
	\end{center}
	\caption{The effective potential for test particles with angular momentum, plotted for $m=1$, $L=1$, $Q=1$,  $q=0.5$ and $\lambda = 10$. The turning points are determined by the intersection of $E$ and the effective potential (i.e. $E_t = V(r_t)$). These include the radius of unstable circular orbits $r_U$, and two other points, $r_S$ and  $r_F$.} 
	\label{fig:M_effectivePotential_0}
\end{figure}

In this section we focus on analyzing the trajectories followed by charged particles with non-zero angular momentum ($L\neq0$). The effective potential in Eq.~\eqref{mpc3.1} has been plotted in Fig.~\ref{fig:M_effectivePotential_0}, in which the turning points $r_t$ correspond to the values of $E=E_t$ that satisfy $E_t = V(r_t)$. The significance of these points is that they do reveal the possible orbits of the test particles. In fact, according to Fig.~\ref{fig:M_effectivePotential_0}, three turning points are highlighted; $r_t = r_U$ (the radius of unstable circular orbits), $ r_t = r_S$ (the distance from the point of scattering) and $r_t = r_F$ (the point of no return, or the capturing distance).
%\begin{figure}[t]
%	\begin{center}
%		\includegraphics[width=9cm]{F1.pdf}
%	\end{center}
%	\caption{Evolution of the effective potential for charged %test particles as a function of the radial coordinate $r$ for %three values of the angular momentum $L=0$ (dashed curve), %$L=0.5$ (meddle curve), and $L=1$ (upper curve). Each curve was %plotted with  $Q=1$ and $\lambda=10$.}
%	\label{F1Potential}
%\end{figure} 

%%%%%%%%%%%%%%%%%%%%%%%%%%%%%%%%%
\subsection{Unstable circular orbits}\label{subsec:circular}

{As observed from the effective potential in Fig.~\ref{fig:M_effectivePotential_0}, the orbits become unstable at the a maximum, whose corresponding radius is limited from above to a circle of radius $r_U$,  where the gravitational attraction caused by the mass of the source, is completely replaced by the
cosmological repulsion caused by the term $\tilde{\varepsilon}$. This radius of unstable circular orbits, or the \textit{static radius} as termed in Refs.~\cite{Stuchlik:1983,Stuchlik:1999}}, is given by the condition $V'(r)\equiv\left.\frac{\partial V(r)}{\partial r} \right|_{r_U} = 0$. Hence, from Eq.~\eqref{mpc3.1} we get
\begin{equation}\label{eq:Vprime=0}
\left.\left( \sqrt{\frac{G(r; L)}{B(r)}} \,\frac{B'(r)}{2}-\sqrt{\frac{B(r)}{G(r; L)}} {L^2\over r^3}
 -\frac{q Q}{\sqrt{2} r^2}\right)\right|_{r_{U}}=0,  \end{equation}
in which the function $G(r; L)$ is defined as
\begin{equation}
    \label{fctg} G(r; L)=m^2 +\frac{L^2}{r^{2}}.
\end{equation}
In fact, the left hand side of Eq.~\eqref{eq:Vprime=0} leads to an incomplete polynomial of twelfth degree in $r$, and hence, it can be solved only numerically. It is however still possible to calculate the proper ($T_\tau$) and the coordinate ($T_t$) periods of these orbits. Combining Eqs.~\eqref{veltau}, \eqref{mpc5} and \eqref{velphi}, and the fact that for a complete orbit $\Delta\phi_U = 2\pi$, we have
\begin{eqnarray}\label{eq:periods_definition}
     T_\tau&\equiv&\Delta\tau = \frac{2\pi r_U^2}{L_U},\label{eq:periodProper}\\
    T_t&\equiv&\Delta t =  \frac{2\pi r_U^2 }{L_U}\,\frac{E_U-V_q(r_U )}{B(r_U)}=T_{\tau}\sqrt{\frac{G_U }{B_U}},\label{eq:periodCoordinate}
\end{eqnarray}
where $G_U\equiv G(r_U; L_U)$ and $B_U\equiv B(r_U)$. Solving Eq.~\eqref{eq:Vprime=0}, we then obtain an expression for $L_U$ as (appendix \ref{app:A0})
\begin{equation}
L_{U}=  \sqrt{\frac{ \mathfrak{b}-\sqrt{ \mathfrak{b}^{2}-4 \mathfrak{a} \mathfrak{c}}}{2 \mathfrak{a}}},
\label{T.25}
\end{equation}
as the angular momentum for the circular orbits, where
\begin{subequations}
\begin{align}
&\mathfrak{a}={(Q^2-2r_U^2)^2\over r_U^6 },
\\
&\mathfrak{b}={2Q^2(1+q^2)\over r_U^2}-{Q^4(2+q^2)\over 2r_U^4}-{8r_U^2-2Q^2(1-q^2)\over \lambda^4},
\\
&\mathfrak{c} ={Q^4(1+2q^2)\over 4r_U^2}-2q^2Q^2+{4r_U^6\over \lambda^4}-{2Q^2r_U^2(1-q^2) \over \lambda^2}.
\end{align}
\end{subequations}
Accordingly, one can obtain the proper frequency
\begin{equation}\label{omegatau}
    \omega_{\tau}= \frac{2\pi}{T_\tau} = \sqrt{\frac{ \mathfrak{b}-\sqrt{ \mathfrak{b}^{2}-4 \mathfrak{a} \mathfrak{c}}}{2 \mathfrak{a} r_U^4}},
\end{equation}
straightly from Eqs. (\ref{eq:periodProper}) and (\ref{T.25}). The coordinate frequency can then be given by the ratio
\begin{equation}\label{p1}
      \frac{\omega_\tau}{\omega_{t}} = \sqrt{\frac{G_U }{B_U}}.
\end{equation}
These values correspond to the velocity of particles on a surface, where they can maintain a circular orbit before falling into the event horizon or escape from it. In the study of particle trajectories, the critical orbits can locate the innermost possible stable orbits around black holes and therefore are of great importance. The test particles, however, can also be scattered at the turning point $r_S$, pursue a hyperbolic motion and escape the black hole. For electrically charged particles, this corresponds to the so-called Rutherford scattering. We continue our discussion by analyzing this kind of orbit.

%%%%%%%%%%%%%%%%%%%%%%%%%%%%%%%%%%
\subsection{Orbits of the first kind and the gravitational Rutherford scattering}

The particle deflection by the CWBH happens when the condition $E_{+}<E < E_U$ is satisfied. This indeed results in two points of approach, $r_t = r_S$ and $r_t = r_F$, at which, $\frac{\mathrm{d}r}{\mathrm{d}\phi}|_{r_t} = 0$ or $E_t= V(r_t)$ (see Fig.~\ref{fig:M_effectivePotential_0}). The relevant equation of motion can be derived from Eqs.~\eqref{veltau} and \eqref{velphi}, giving
\begin{equation}\label{tl11}
\left(\frac{\mathrm{d}r}{\mathrm{d}\phi}\right)^2=\frac{\mathcal{P}(r)}{\upsilon^2}, 
\end{equation}
where
\begin{subequations}
\begin{align}
    &\mathcal{P}(r)\equiv r^6+\mathcal{A}  r^4 + \mathcal{B} r^3+\mathcal{C} r^2+\mathcal{D},\label{pol6}\\
    &\upsilon=\frac{L \lambda}{m},\label{ups}
 \end{align}
\end{subequations}
with
\begin{subequations}\label{eq:ABCD}
\begin{align}
  & \mathcal{A}=\upsilon^2 \left( \frac{E^2-m^2}{L^2}+{1 \over \lambda^2}\right),\\
    &  \mathcal{B}=-2 E \upsilon^2 \left({q Q \over \sqrt{2} L^2}\right),\\
   & \mathcal{C}=\frac{\mathcal{D}}{L^2}-\frac{\mathcal{B}}{2 E}-\upsilon^2,\\
 &  \mathcal{D}=\upsilon^2 \left({m Q\over 2}\right)^2.
   \end{align}
\end{subequations}
To determine the turning points $r_S$ and $r_F$, one therefore needs to solve $\mathcal{P}(r_t) = 0$, which is an incomplete equation of sixth degree in $r$, and values of $r(\phi)$ can therefore be obtained through numerical methods. To deal with this problem, we pursue the inverse process and find an analytical expression for $\phi(r)$. The behavior of $r(\phi)$ can then be demonstrated by means of numerical interpolations.

\begin{figure}[t]
	\begin{center}
		\includegraphics[width=8cm]{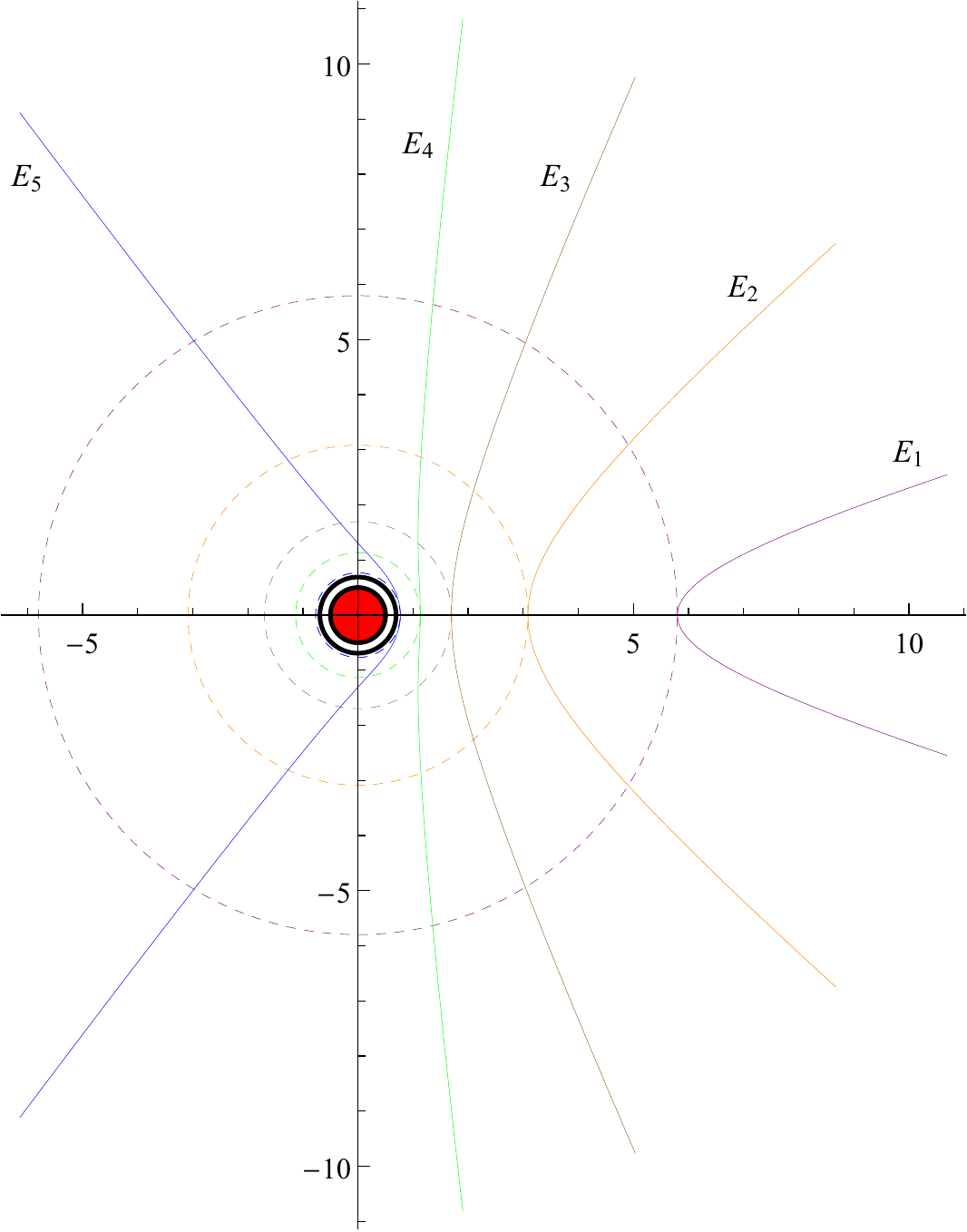}
	\end{center}
	\caption{The Rutherford scattering plotted for $m=1$, $Q=1$, $q=0.5$, $\lambda=10$ and $L=1$. For these values, $r_+ = 0.50$, $r_U=0.69$ and $E_U = 1.72$. The trajectories have been plotted for $E_1 = 0.88$, $E_2=1.1$, $E_3=1.3$, $E_4=1.5$ and $E_5=1.7$,  while their corresponding scattering distance ($r_S$) have been indicated by dashed circles. As it is observed, the condition $E_5\approx E_U$ has made the corresponding shape of the scattering to be of a convex form, showing an appeal to the critical orbits.}
	\label{fig:scattering}
\end{figure}

To proceed with this method, let us consider that $\mathcal{P}(r)$ has two distinct real roots, corresponding to the turning points $r_1=r_S$  and $r_2=r_F$, two equal and negative real roots, say $r_3=r_4<0$ , and finally, a complex conjugate pair $r_5$  and $r_6=r_5^{*}$. Accordingly, we can recast $\mathcal{P}(r)$ as
\begin{eqnarray}\label{Pr6}
\mathcal{P}(r) &=& \prod_{j=1}^{6}(r-r_{j})\nonumber\\
& =& (r-r_S)(r-r_F)(r-r_5)(r-r_3)^2(r-r_5^*).
\end{eqnarray}
{Taking into account the outgoing trajectories, the equation of motion \eqref{tl11} can then be written as}
 \begin{equation}\label{phi1}
 \phi(r) =\upsilon \int_{r_S}^{r}{\ed r\over  \sqrt{\mathcal{P}(r)} }.
 \end{equation}
Particles reaching $r_S$, experience a hyperbolic motion around the black hole and then escape to infinity. This kind of motion, known as orbit of the first kind (OFK) \cite{Chandrasekhar:579245,cov04}, has the significance of gravitational Rutherford scattering when the test particles are electrically charged. Considering the change of variable
\begin{equation}\label{eq:cv1}
	u_j \doteq {1\over (r_j/r_S)-1},~~~ j=\{2,3,5,6\},
\end{equation}
the above integral results in (appendix \ref{app:angular})
  \begin{equation}\label{phi1c}
 \phi(r) =\kappa_0\left[\ss(U)-{u_3\over 4}\mathfrak{F}(U)\right],
 \end{equation}
 where $\ss(x)\equiv \ss(x, \mathbf{g}_2, \mathbf{g}_3)$ is the inverse Weierstra$\ss$ $\wp$ function, and
\begin{equation}
\mathfrak{F}(U)=\frac{1}{\wp'(\Omega_S)}
\left[2\zeta(\Omega_S) \ss(U)
+\ln\left|\frac{\sigma\left(\ss(U)-\Omega_S\right)}
{\sigma\left(\ss(U)+\Omega_S\right)}\right|
\right],
\end{equation}
also contains the Weierstra$\ss$ Zeta and Sigma functions ($\zeta(y)$ and $\sigma(y)$,
respectively)\footnote{By definition, we have \cite{handbookElliptic}
\begin{equation*}
    \ss(x) \equiv y = \int_x^\infty\frac{\ed \mathfrak{t}}{\sqrt{4\mathfrak{t}^3-g_2 \mathfrak{t}-g_3}}.
\end{equation*}
Then the inverse function $x = \wp(y,g_2,g_3)\equiv\wp(y)$ defines the elliptic Weierstra$\ss$ $\wp$ function with the coefficients $g_2$ and $g_3$, for which  
\begin{equation*}\label{eq:diff-wp-def}
\wp'(y)\equiv\frac{\mathrm{d}}{\mathrm{d}y}\wp(y) = - \sqrt{4\wp^3(y)-g_2\wp(y)-g_3}.
\end{equation*}
The two other related functions, namely the Weierstra$\ss$ Zeta and Sigma functions, are defined as
\begin{eqnarray*}
 \zeta(y) &=& -\int\wp(y)\ed y,\\
 \sigma(y) &=& e^{\int\zeta(y)\ed y}.
\end{eqnarray*}}.
Here, we have defined
\begin{subequations}\label{radialg2g3}
\begin{align}
&U \equiv U(r) = {1\over 4\left( \frac{r}{r_S} -1\right)}+{\mathbf{a}\over 12},\\
&\Omega_S = \ss\,\left( {\mathbf{a}\over 12}-{1\over 4\left( \frac{r_3}{r_S}-1\right)}\right),\\
&\mathbf{g}_2 = {\mathbf{a}^2\over 12}-{\mathbf{b}\over  4},\\
&\mathbf{g}_3 = \frac{1}{16}\left({\mathbf{a}\mathbf{b}\over 3}- {2\mathbf{a}^3\over 27}-\mathbf{c}
\right),
\end{align}
\end{subequations}
with
\begin{subequations}
\begin{align}\label{abc}
&\mathbf{a} = u_2+u_5+u_6,\\
&\mathbf{b} = u_2(u_5+u_6)+u_5 u_6,\\ 
&\mathbf{c} = u_2 u_5 u_6.
\end{align}
\end{subequations}
The \textit{scattering angle} in Eq.~\eqref{phi1c} gives the change in the particles' orientation as they approach and recede the black hole at the scattering point $r_S$. To illustrate their corresponding trajectories, we make a list of points $\left(r_t,\phi(r_t)\right)$ and then find the numerical interpolating function of $r(\phi)$. The resultant OFK trajectories have been illustrated in Fig.~\ref{fig:scattering} for particles of different values for $E$. As it is observed, the scattering can be formed convexly (approaching) or concavely (receding).
For particles coming from infinity, the scattering angle can be written as \cite{Villanueva:2015,Fathi:2020b}
\begin{equation}\label{eq:scatterin_angle0}
    \vartheta = 2\phi_\infty - \pi,
\end{equation}
 in which $\phi_\infty \equiv \phi(\infty)$. Accordingly, from Eq.~\eqref{phi1c} we have
 \begin{equation}\label{eq:scatterin_angle1}
     \vartheta = -\pi + 2\kappa_0 \left[
     \ss\left(\frac{\mathbf{a}}{12}\right)+\frac{u_3}{4}\left\{
     \frac{1}{\sqrt{\frac{1}{432} \left(
     4\mathbf{a^3}-18\mathbf{a}\mathbf{b}+27 u_3\left(
    \mathbf{b}-\mathbf{a} u_3 + u_3^2\right)\right)}}
     \left[
        2\zeta(\Omega_S) \ss\left(\frac{\mathbf{a}}{12}\right)
        +\ln\left|
        \frac{\sigma\left(\ss\left(\frac{\mathbf{a}}{12}\right)-\Omega_S\right)}{\sigma\left(\ss\left(\frac{\mathbf{a}}{12}\right)+\Omega_S\right)}
        \right|
        \right]
        \right\}
        \right].
 \end{equation}
The value of $\vartheta$ is specified directly by the initial $E$ and the corresponding particular solutions $r_j$, which are determined by the equation $E=V(r)$. These values therefore, cannot be considered to evolve in terms of a single variable. However, one can calculate the scattering angle for each particular trajectory, by applying Eq.~\eqref{eq:scatterin_angle1}. Additionally, the differential angular range of the scattered particles at the angle $\vartheta$, is given by the solid angle element $\ed\rO = \sin\vartheta \ed\vartheta \ed\phi$. In this regard, and defining the impact parameter $b=\frac{L}{E}$, the cross-sectional area of the scattered particles has the differential form $\ed\Sigma = b\, \ed\phi\, \ed b$ \cite{Fathi:2020b}. Therefore, the differential cross section of the scattering is given by
\begin{equation}\label{eq:diffCross-def}
    \Sigma(\vartheta) \doteq \frac{\mathrm{d}\Sigma}{\mathrm{d}\rO} = \frac{b}{\sin\vartheta} \left|\frac{\partial b}{\partial\vartheta}\right|.
\end{equation}
From Eqs.~\eqref{phi1c} and \eqref{eq:scatterin_angle0} we have
\begin{equation}\label{eq:Theta_recast}
    \frac{1}{2\kappa_0}(\vartheta+\pi)=\varphi_{1} + \varphi_{2},
\end{equation}
in which
\begin{subequations}\label{eq:var1.var2}
\begin{align}
   & \varphi_{1} \equiv\ss\left( {\mathbf{a}\over 12}\right),\\ 
   &  \varphi_{2} \equiv -{u_3\over 4}\mathfrak{F}\left( {\mathbf{a}\over 12}\right).
\end{align}
\end{subequations}
We define
\begin{equation}\label{eq:Psi}
    \Psi(L) \doteq \wp\left(
    \frac{\vartheta+\pi}{2\kappa_0}
    \right) = \wp\left(
    \varphi_{1} + \varphi_{2}
    \right),
\end{equation}
or \cite{handbookElliptic}
\begin{equation}\label{eq:Psi(L)}
    \Psi(L) = \frac{1}{4}\left[
    \frac{\wp'(\varphi_{1})-\wp'(\varphi_{2})}{\wp(\varphi_{1})-\wp(\varphi_{2})}
    \right]^2-\wp(\varphi_{1})-\wp(\varphi_{2}).
\end{equation}
Now, applying the definition in Eq.~\eqref{eq:Psi(L)}, we can recast Eq.~\eqref{eq:diffCross-def} as
\begin{equation}\label{eq:diffCross-def_1}
    \Sigma(\vartheta) = b\csc\vartheta\left|
    \frac{\partial\Psi}{\partial\vartheta}
    \right|
    \left|
    \frac{\partial b}{\partial\Psi}
    \right|
    = \frac{1}{4\kappa_0}\csc\vartheta\left|
    \wp'\left(
    \frac{\vartheta+\pi}{2\kappa_0}
    \right)
    \right|
    \left|
    \frac{\partial b^2}{\partial\Psi}
    \right|,
\end{equation}
for which, the identity $\frac{\partial b^2}{\partial\Psi} = \frac{\partial b^2/\partial L}{\partial\Psi/\partial L}$ yields
\begin{equation}\label{eq:sigmaTheta_2}
    \Sigma(\vartheta) = \frac{ L}{2\kappa_0E^2}\csc\vartheta\left|
    \wp'\left(
    \frac{\vartheta+\pi}{2\kappa_0}
    \right)
    \right|
    \left|
    \frac{\partial\Psi}{\partial L}
    \right|^{-1}.
\end{equation}
The expression of $\Psi$ is analytically complicated. However, as before, the value of Eq.~\eqref{eq:sigmaTheta_2} can be numerically calculated regarding definite initial values for distinct scattered trajectories.\\

In this section, we analyzed the angular trajectories of infalling particles and demonstrated the corresponding Rutherford scattering which is the most significant feature of such trajectories. However, scattering can happen, as well, in the absence of the particles' angular momentum, which is what we deal with in the next section by studying the radial trajectories of the  test particles.

%%%%%%%%%%%%%%%%%%%%%%%%%%%%%%%%%%%%%%%%%%%%%%%%%%%%%%%
\section{Radial Trajectories}\label{sec:radial}

To discuss the horizon crossing process, as viewed by the comoving and the distant observers, it is common to use a framework, in which, the particles fall radially onto the black hole, without possessing any angular momentum. As well as in the case of angular orbits, purely radial motion can itself be ramified into several kinds. Further in this section, these kinds of motion are discussed regarding the horizon crossing of the comoving observers and the \textit{frozen} particles as viewed by distant observers (see Refs.~\cite{Zeldovich:2014,Ryder:2009} for the notion of a frozen infalling object onto a black hole).

\begin{figure}[t]
	\begin{center}
		\includegraphics[width=8.5cm]{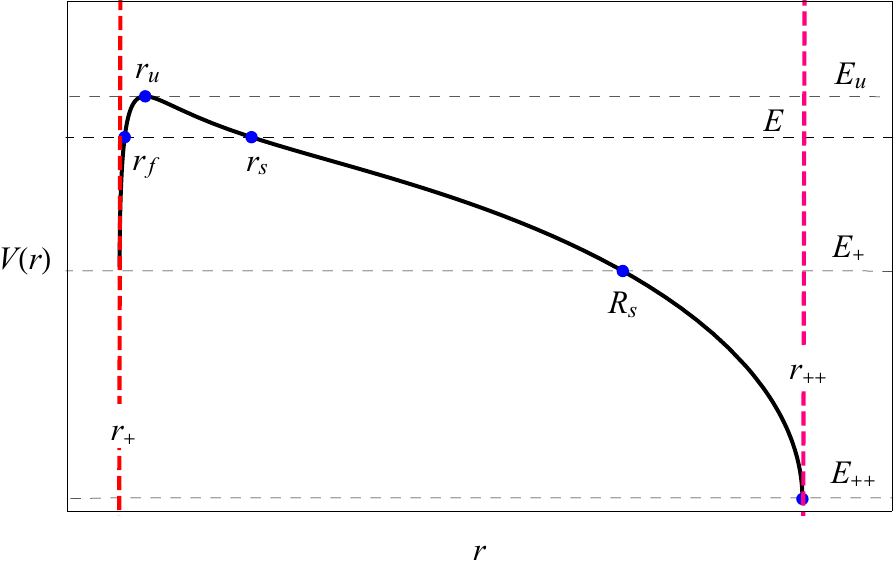}
	\end{center}
	\caption{The effective potential for radially moving particles plotted for $m=1$, $Q=1$, $q=0.5$ and $\lambda=10$. The maximum distance of unstable motion, $r_u$, and the two turning points $r_s$ and $r_f$ have been indicated in accordance with their corresponding values of $E$. In particular, the point $R_s$ is related to the distance at which the particles of the constant of motion $E_+$, experience their Rutherford scattering. }
	\label{fig1}
\end{figure}

The vanishing angular momentum of the radially moving particles, reduces the effective potential in Eq.~\eqref{mpc3.1} to
\begin{equation}\label{rt1} 
V_r(r)= V_q(r)+m\sqrt{B(r)},
\end{equation}
whose behavior has been plotted in Fig.~\ref{fig1}.
Accordingly, the motion becomes unstable where $V'_r(r)=0$, solving which, leads to the maximum distance of the unstable motion, reading as 
\begin{equation}\label{rt1} 
r_u=\left[\tilde{ \alpha}-\sqrt{\tilde{ \alpha}^2-\tilde{\beta}}\right]^{1/2},
\end{equation} 
where (see appendix \ref{app:Ap})
\begin{subequations}\label{eq:tildealphatildebeta}
\begin{align}
  &\tilde{\alpha}=\sqrt{\tilde{U}-{\tilde{a}\over 6}}, \\
&\tilde{\beta}=2\tilde{\alpha}^2 +{\tilde{a}\over 2}+{\tilde{b}\over 4\tilde{\alpha}}  , 
\end{align}
\end{subequations}
given that
\begin{equation}\label{eq:tildeU}
\tilde{U}=2\sqrt{{\tilde{\eta}_2\over 3}} \cosh\left(\frac{1}{3}\arccosh\left({3\over 2}\tilde{\eta}_3\sqrt{{3\over \tilde{\eta}_2^3}}  \right) \right),
\end{equation}
with
\begin{subequations}\label{eq:tildeabctildeeta}
\begin{align}
   &\tilde{a}=-{Q^2\lambda^2\over 2}\left(1- {q^2\over m^2}\right),\\
   &\tilde{b}=-{q^2Q^2\lambda^4\over 2m^2},\\
   &\tilde{c}={Q^4\lambda^4\over 16}\left(1+{2q^2\over m^2}\right),\\
   &\tilde{\eta}_2={\tilde{a}^2\over 48}+{\tilde{c}\over 4},\\
   &\tilde{\eta}_3={\tilde{a}^3\over 864}+{\tilde{b}^2\over 64}-{\tilde{a} \tilde{c}\over 24}.
\end{align}
\end{subequations}
Taking into account $E_u \equiv V_r(r_u)$, as in the angular case, possible motions are categorized based on the value of $E$ compared with its critical value, $E_u$:
 \begin{itemize}
	\item \textit{Frontal Rutherford scattering of the first and the second kinds (RSFK and RSSK)}: For 	$E_{++}<E<E_u$, the potential allows for a turning point $r_s$ ($r_u<r_s<r_{++}$) which corresponds to the scattering distance (RSFK). In the case that $E_{+}<E<E_u$, there is also another turning point $r_f$ ($r_+<r_f<r_{u}$), from which, the trajectories are captured into the event horizon (RSSK).
	
	\item \textit{Critical radial motion}: For $E=E_u$, the particles can stay on an unstable radial distance of radius $r=r_u$.  Therefore, those coming from the initial distances $r_i$ or $d_i$  ($r_u < r_i < r_{++}$ and $r_+ < d_i < r_u$ respectively), will ultimately fall on $r_u$, or cross the horizons.
	
%	\item \textit{Radial capture}: If $E>E_u$, particles coming from a finite distance $\rho_0$ ($r_+ < \rho_0 < r_{++}$), are pulled towards the horizons from the same distance. 
\end{itemize} 
Now, let us rewrite the radial velocity relations, given in Eqs.~\eqref{veltau} and \eqref{mpc5}, as
\begin{eqnarray}
\left(\frac{{\rm d}r}{{\rm d}\tau}\right)^{2}&=&\frac{m^2 \mathfrak{p}(r)}{\lambda^2 r^2},
\label{vel1}\\
\left(\frac{{\rm d}r}{{\rm d}t}\right)^{2}&=&\frac{m^2
(r^2-r^2_{+})^2(r^2_{++}-r^2)^2 \mathfrak{p}(r)}{E^2 \lambda^6 r^4(r-\frac{\sqrt{2} q Q}{E})^2},\label{vel2}
\end{eqnarray}
with 
\begin{equation}\label{eq:mathfrakp}
    \mathfrak{p}(r)\equiv  r^4+\bar{a} r^2 +\bar{b} r+\bar{c},
\end{equation}
where
\begin{subequations}\label{eq:barabarbbarc}
\begin{align}
&\bar{a}={(E^2-m^2) \lambda^2\over m^2},\\
&\bar{b}=-{\sqrt{2}qQE \lambda^2\over m^2},\\
&\bar{c}=\frac{Q^2(m^2+2q^2)\lambda^2}{4\,m^2}.
\end{align}
\end{subequations}
Applying the above equations, in this section, the radial behavior of the time parameter for the moving particles is calculated, together with that for the distant observers.

%%%%%%%%% frontal scattering
\subsection{Frontal  scattering}

\begin{figure}[t]
	\begin{center}
		\includegraphics[width=8.5cm]{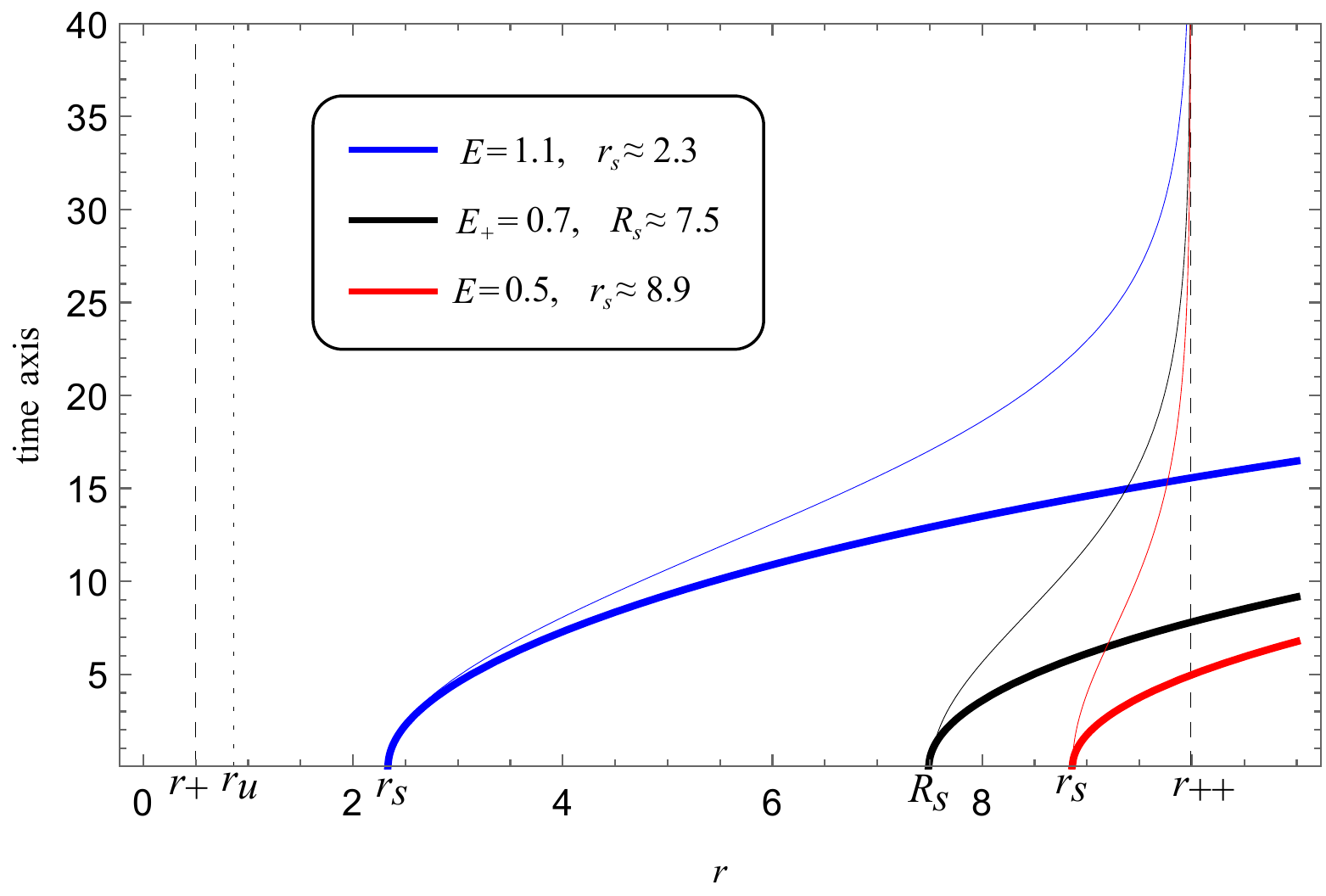}
	\end{center}
	\caption{The radial behavior of the proper and coordinate times in the RSFK, for three scattering points and their corresponding values of $E$. After the scattering, the comoving observers (thick line) see a horizon crossing. This is while a distant observer (thin line) never observes this (frozen falling particles). The plots have been done for $m=1$, $Q=1$, $q=0.5$ and $\lambda=10$. }
	\label{fig:RSFK}
\end{figure}

As it is inferred from the effective potential in Fig.~\ref{fig1}, particles can encounter two turning points $r_s$ and $r_f$ which are located at either sides of the critical distance ($r_f < r_u <r_s$). These turning points do lead the trajectories to different fates. Particles with $E_{++}<E<E_+$, however, can only escape the black hole by being scattered at the only possible turning point $r_s$. Same as discussed in Sec.~\ref{sec:angular}, the turning points are where the particles' coordinate velocity vanishes, which for the radial trajectories requires $\mathfrak{p}(r) = 0$ in Eq.~\eqref{eq:mathfrakp}, giving
\begin{eqnarray}\label{rs} 
r_s&=&\bar{\alpha}+\sqrt{ \bar{\alpha}^2-\bar{\beta}}, \\
r_f&=&\bar{\alpha}-\sqrt{ \bar{\alpha}^2-\bar{\beta}}. 
\end{eqnarray}
These radii are basically based on the same components as in Eqs.~\eqref{eq:tildealphatildebeta}--\eqref{eq:tildeabctildeeta}, and we only need to replace $\tilde{a}\rightarrow\bar{a}$, $\tilde{b}\rightarrow\bar{b}$ and $\tilde{c}\rightarrow\bar{c}$, according to the values given in Eqs.~\eqref{eq:barabarbbarc}.
Having determined the turning points, the polynomial $\mathfrak{p}(r)$ can be decomposed accordingly. As described above, the first kind scattering (RSFK) happens when the particles approach at $r_s$, which is now considered as their initial position. Therefore Eq.~\eqref{vel1} can be solved as (appendix \ref{app:A1})
\begin{equation}\label{eq:tau(r)-int-4}
\tau (r)={-\lambda \over m\sqrt{\gamma_0}}\left[ 
\ss(\mathrm{U})+
{1\over 4} \mathrm{F}(\mathrm{U})
\right], 
\end{equation}
where 
\begin{equation}\label{eq:F(U)}
	\mathrm{F}(\mathrm{U})=\frac{1}{\wp'(\Omega_s)}
	\left[2\zeta(\Omega_s)\ss(\mathrm{U})
	+\ln\left|\frac{\sigma\left(\ss(\mathrm{U})-\Omega_s\right)}
	{\sigma\left(\ss(\mathrm{U})+\Omega_s\right)}\right|
	\right],
\end{equation}
and the function $\mathrm{U}(r)$ and  the Weierstra$\ss$ coefficients are given as
\begin{subequations}\label{radialg2g3}
\begin{align}
	& \mathrm{U}(r)={r_s \over 4(r-r_s)}+{\gamma_1\over 12 \gamma_0},\\
	& \Omega_s =\ss\left({\gamma_1\over 12 \gamma_0}\right),\\
	& \bar{g}_2 = {\gamma_1^2\over 12 \gamma_0^2}-{1\over  \gamma_0},\\
	&\bar{g}_3 =\frac{1}{16}\left({4\gamma_1\over 3 \gamma_0^2}- {2\gamma_1^3\over 27 \gamma_0^3}-{1\over  \gamma_0}
	\right),
\end{align}
\end{subequations}
with 
\begin{subequations}\label{eq:gamma12}
\begin{align}
 &\gamma_1=6+\frac{\bar{a}}{r_s^2},\\
 &\gamma_0=4+\frac{2\bar{a}}{r_s^2}+\frac{\bar{b}}{r_s^3}.
\end{align}
\end{subequations}
The relation in Eq.~\eqref{eq:tau(r)-int-4} measures the radial change of the time parameter for observers comoving with the particles. For distant observers, such measurement is done on the coordinate time, whose evolution can be obtained by exploiting the velocity in Eq.~\eqref{vel2}. Applying the same method as before, we obtain 
\begin{equation}\label{teradsct}
t(r)=-\delta_0 \left[\ss(\mathrm{U})+
{1\over 4}\sum_{k=1}^{4} \delta_k \mathrm{F}_k(\mathrm{U})\right],
\end{equation}
where
\begin{equation}\label{eq:Fj(U)}
 F_k(\mathrm{U}) = \frac{1}{\wp'(\Omega_k)}\left[2\zeta(\Omega_k)\ss(\mathrm{U})
+\ln\left|\frac{\sigma\left(\ss(\mathrm{U})-\Omega_k\right)}
{\sigma\left(\ss(\mathrm{U})+\Omega_k\right)}\right|
\right],
\end{equation}
and
\begin{equation}\label{eq:Omegaj}
    \Omega_k= \ss\left({\gamma_1\over 12\, \gamma_0}+{z_k\over 4} \right),
\end{equation}
in which $z_k \doteq {1 \over (r_k/r_s)-1}$, with $r_1\equiv r_+$, $r_2\equiv -r_{+}$,  $r_3\equiv r_{++}$ and $r_4\equiv -r_{++}$, and the coefficients are expressed as
\begin{subequations}
\begin{align}
\delta_{0} &= {\lambda^2\,E\over m\sqrt{\gamma_0}} {z_1z_2z_3z_4\over z_5r_s^2},\\
\delta_{1} &=  {(z_1+1)^2z_1(z_1-z_5)\over (z_1-z_2)(z_1-z_3)(z_1-z_4)},\\
\delta_{2} &=   {(z_2+1)^2z_2(z_2-z_5)\over (z_2-z_1)(z_2-z_3)(z_2-z_4)},\\
\delta_{3} &=  {(z_3+1)^2z_3(z_3-z_5)\over (z_3-z_1)(z_3-z_2)(z_3-z_4)},\\
\delta_{4} &=  {(z_4+1)^2z_4(z_4-z_5)\over (z_4-z_1)(z_4-z_2)(z_4-z_3)}.
\end{align}
\end{subequations}
Since these trajectories escape the black hole, they will eventually confront the cosmological horizon. In Fig.~\ref{fig:RSFK}, the temporal relations in Eqs.~\eqref{eq:tau(r)-int-4} and \eqref{teradsct} have been used to demonstrate the RSFK, as observed by comoving and distant observers, for three different scattering distances. As it is seen, for the comoving observers, a horizon crossing occurs within a finite time after the particles are scattered at $r_s$. This is while for the distant observers it takes an infinite amount time for the particles to pass the horizon; in other words, they appear frozen.

%%%%%%%%%%%%%%%%%%%%%% RSSK
\subsection{Frontal scattering of the second kind}

\begin{figure}[t]
	\begin{center}
		\includegraphics[width=8.5cm]{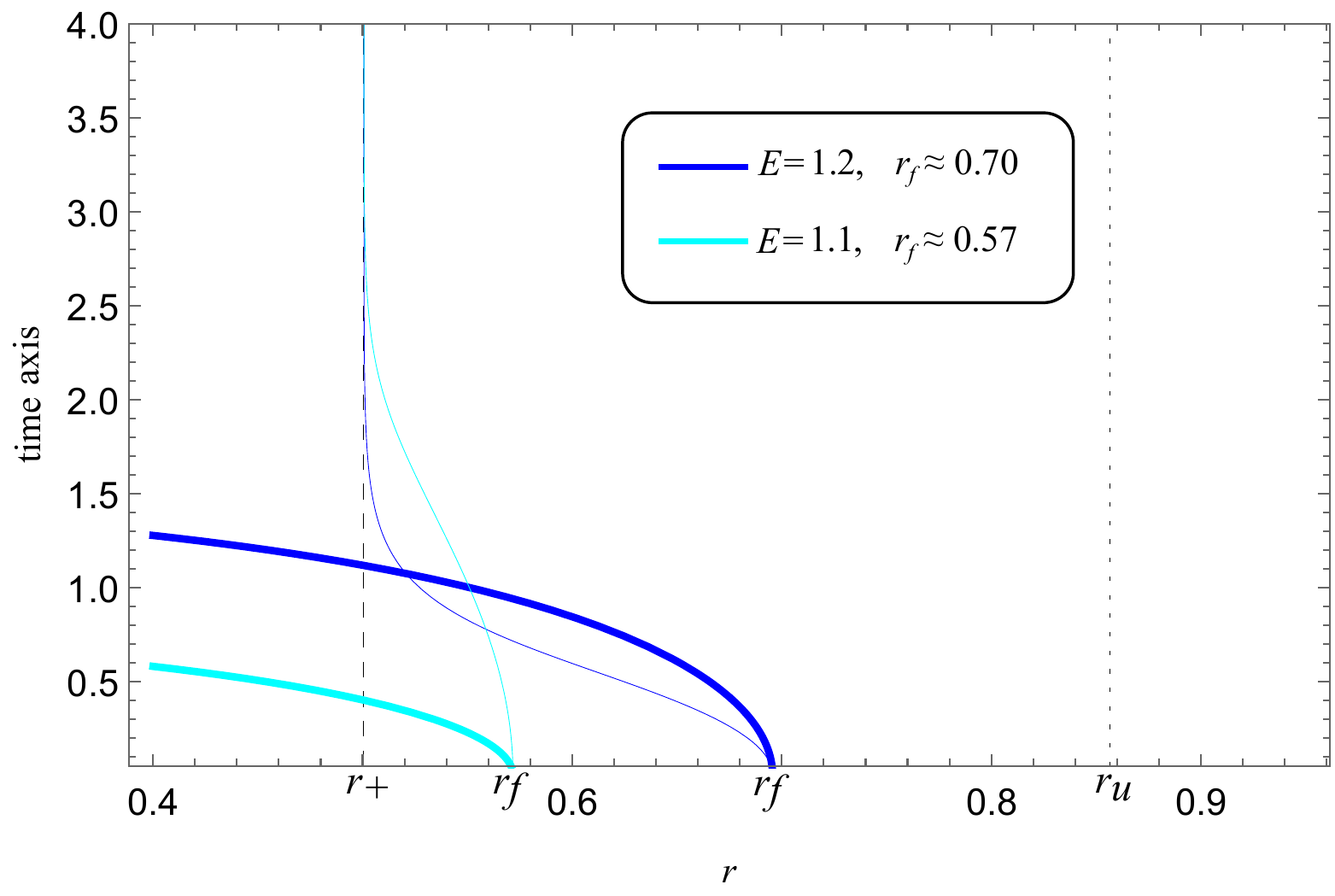}
	\end{center}
	\caption{The RSSK for two different scattering points and their corresponding values of $E$, plotted for $m=1$, $Q=1$, $q=0.5$ and $\lambda=10$. }
	\label{fig:RSSK}
\end{figure}

By switching the scattering distance to $r_f$, the particles experience the RSSK and they confront the event horizon. The corresponding equations of motion are the same as those in the case of RSFK and are given by exchanging $r_s\rightarrow r_f$ in the relations. The corresponding temporal parameters have been demonstrated in Fig.~\ref{fig:RSSK}. Same as before, the comoving and distance observers see different fates for the infalling particles, but here, regarding the capture process by the event horizon.

%%%%%%%%%%%%%%%%%%%%%%%%%%% critical motion
\subsection{Critical radial motion}

\begin{figure}[t]
	\begin{center}
		\includegraphics[width=8.2cm]{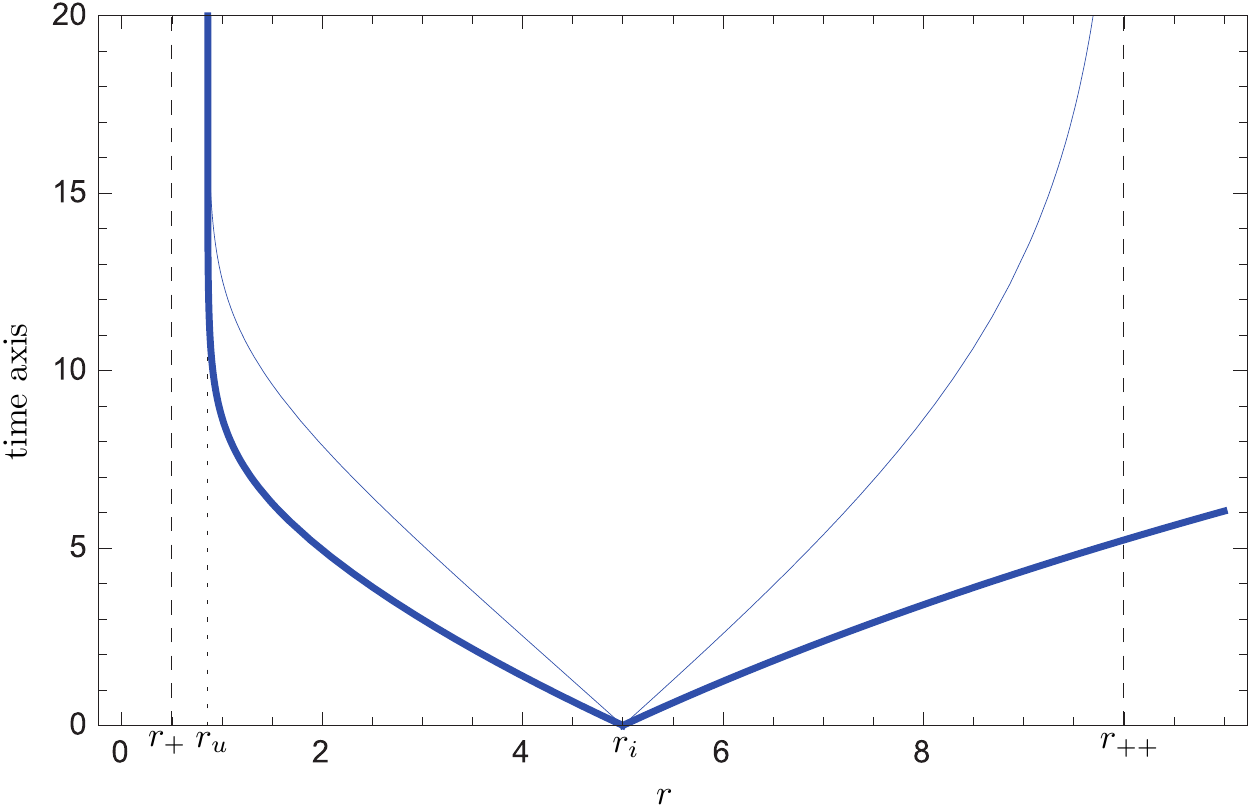}~(a)
		\hfill
			\includegraphics[width=8.2cm]{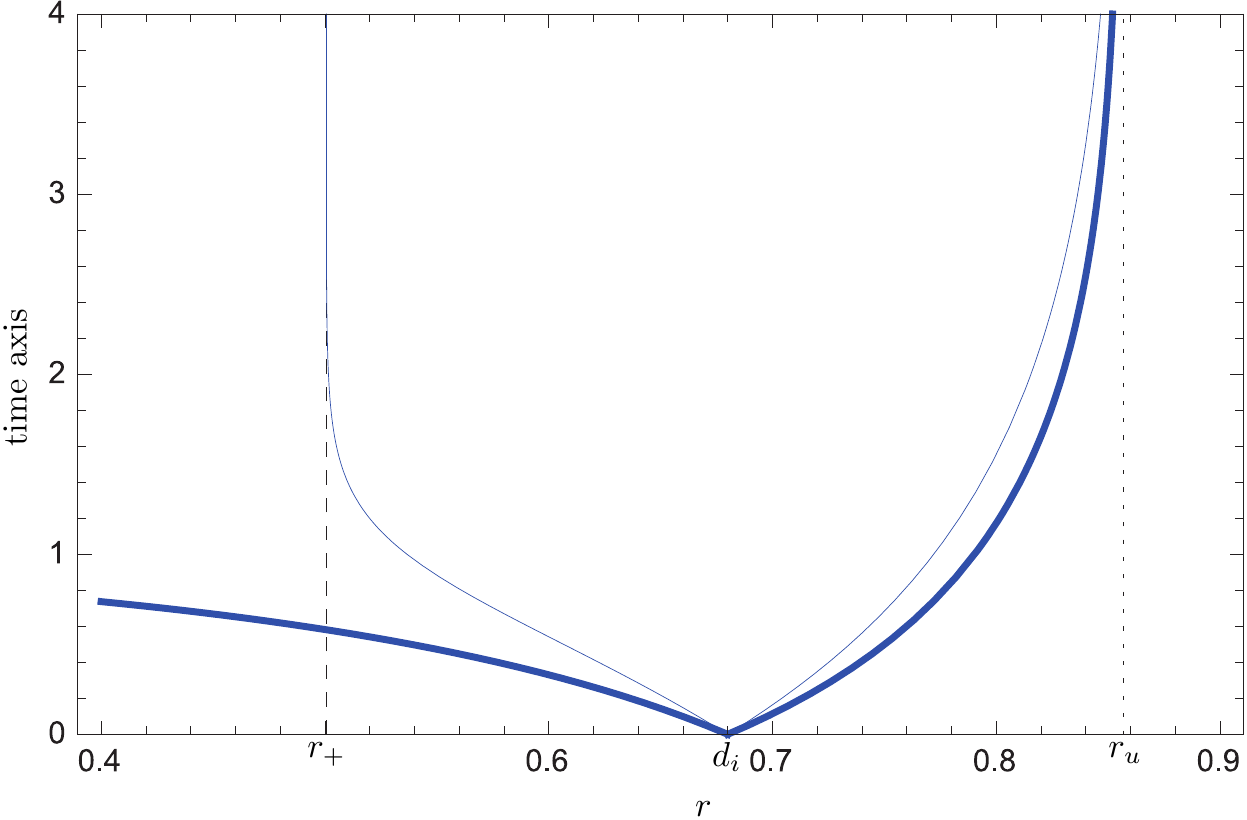}~(b)
	\end{center}
	\caption{The critical radial motion for fates $I$ and $II$, plotted for comoving (thick line) and distant (thin line) observers, by letting $m=1$, $Q=1$, $q=0.5$ and $\lambda=10$. The trajectories have been specified for particles approaching from (\textbf{a}) $r=r_i=5$ and (\textbf{b}) $r=d_i=0.68$.}
	\label{fig:radialCritical}
\end{figure}

In the case that $E = E_u$, the unstable (critical) motion of particles depends on whether they approach from $r_i>r_u$ or from $d_i<r_u$.  According to the discontinuity of $\frac{\mathrm{d}\tau}{\mathrm{d}r}$ and $\frac{\mathrm{d}t}{\mathrm{d}r}$ at $r_i$ and $d_i$, we can expect two different behaviors for the approaching particles, in the sense that they either fall on $r = r_u$ (fate $I$) or be pulled towards the horizons (fate $II$). These are revealed by integrating the equations of motion for the time parameters. For particles coming from $r_i$, we obtain
\begin{eqnarray}\label{mrc2}
\tau_I (r)&=&\pm {\lambda\over m} \left[ \tau_A (r)- \tau_B (r)-\tau_A (r_i)+\tau_B (r_i) \right],  \label{mrc1}\\
\tau_{II} (r)&=&\mp {\lambda\over m} \left[ \tau_A (r)- \tau_B (r)-\tau_A (d_i)+\tau_B (d_i) \right],
\end{eqnarray}
for the comoving observers, where
\begin{subequations}
\begin{align}
&\tau_A(r)=\arcsinh \left({r+r_u\over \sqrt{\bar{a}+2r_u^2}}\right),\\
&\tau_B(r)={r_u\over \sqrt{6r_u^2+\bar{a}}}~\arcsinh \left({6r_u^2+\bar{a}+2r_u(r-r_u)\over |r-r_u|\sqrt{\bar{a}+2r_u^2}}\right).
\end{align}
\end{subequations}
For the distant observers, we get
\begin{eqnarray}\label{mrc1}
t_I (r)&=&\pm {\lambda^3 E\over m r_u^2} \sum_{n=0}^{4}\,\varpi_n\,\left[t_n(r)-t_n(r_i)\right], \\
t_{II} (r)&=&\mp {\lambda^3 E\over m r_u^2} \sum_{n=0}^{4} \varpi_n \left[t_n(r)-t_n(d_i)\right],\label{mrc2}
\end{eqnarray}
where
\begin{subequations}
\begin{align}
&t_n(r)= {r_u\over \sqrt{R_n^2}}~\arcsinh \left({R_n^2+(r_u+r_n)(r-r_n)\over |r-r_u|\sqrt{\bar{a}+2r_u^2}}\right),\\
&R_n^2 = 3r_u^2+\bar{a}+2r_u r_n +r_n ^2,
\end{align}
\end{subequations}
and the coefficients are given as
\begin{subequations}
\begin{align}
\varpi_{0} &= {r_u^3(r_u-r_5)\over (r_1-r_u)(r_2-r_u)(r_3-r_u)(r_4-r_u)},\\
\varpi_{1} &= {r_u r_1^2(r_1-r_5)\over (r_1-r_u)(r_1-r_2)(r_1-r_3)(r_1-r_4)},\\
\varpi_{2} &= {r_u r_2^2(r_2-r_5)\over (r_2-r_u)(r_2-r_1)(r_2-r_3)(r_2-r_4)},\\
\varpi_{3} &= {r_u r_3^2(r_3-r_5)\over (r_3-r_u)(r_3-r_1)(r_3-r_2)(r_3-r_4)},\\
\varpi_{4} &= {r_u r_4^2(r_4-r_5)\over (r_4-r_u)(r_4-r_1)(r_4-r_2)(r_4-r_3)},
\end{align}
\end{subequations}
in which $r_0 \equiv r_u$, $r_1\equiv r_+$, $r_2\equiv -r_{+}$,  $r_3\equiv r_{++}$ and $r_4\equiv -r_{++}$. The critical radial behavior of temporal parameters, as measured by the comoving and the distant observers, have been plotted in Fig.~\ref{fig:radialCritical}, separately for the initial points $r_i$ and $d_i$. In each of the diagrams, the cases $I$ and $II$ have been demonstrated and the horizon crossing is shown accordingly. Once again, particles appear frozen to distant observers as they approach the horizons. \\

So far, both the angular and the radial trajectories of charged particles were studied and possible analytical solutions to the equations of motion were given. It is however worth mentioning that the study of the physical properties of moving particles is not summarized to the evolution of a single particle's trajectory. In the case that a bundle of trajectories is taken into account, definite kinematical parameters will play important roles in the characterization of a \textit{flow} of particle trajectories. Accordingly, and in the next section, we consider such a flow of particles and study how it reacts to the internal and external forces acting on the world-lines.

%%%%%%%%%%%%%%%%%%%%%%%%%%%%%% deviation
\section{A congruence of infalling charged particles}\label{sec:deviation}

In this section, we consider a bundle of particle trajectories, which together, constitute a congruence of world-lines that fall onto the CWBH. Essentially, the congruence kinematics is a tool to
inspect the Penrose–Hawking singularity theorems \cite{Penrose:1965,Hawking:1965,Hawking:1966,Penrose:2002} and is accurately formulated by the well-known Raychaudhuri equation \cite{Raychaudhuri:1955}. This equation formulates the way the congruences would evolve their cross-sectional (transverse) area (for a good review see Ref.~\cite{Kar:2006}).

Here, we switch our discussion to the possibility of applying some geometrical methods in order to demonstrate the deviation of a congruence of time-like trajectories while they pass the black hole. For particles passing a RN black hole, this deviation has been studied in detail in Refs.~\cite{Balakin2000,Heydari-Fard:2019}.  

In the geometric sense, the congruence deviation gives the relative acceleration between the curves that are generated by the tangential vector $\bm u$, in terms of the Jacobi (deviation) vector field $\bm\xi$. This vector field resides on the curves that connect points of equal $\tau$ on smooth planes of world-lines. These vectors satisfy \cite{Wald:1984,Poisson:2004}
\begin{equation}\label{eq:Lie_1}
    {\mathfrak{L}}_{\bm{u}}\bm{\xi}={\mathfrak{L}}_{\bm{\xi}}\bm{u},
\end{equation}
where $\mathfrak{L}_{\bm{X}}$ indicates the Lie differentiation with respect to a vector field $\bm{X}$. The above equation therefore can be recast as
\begin{equation}\label{eq:Lie_2}
    {\xi^\mu}_{;\nu} u^\nu = {u^\mu}_{;\nu}\xi^\nu.
\end{equation}
In above, the semicolons correspond to covariant differentiation. Note that, the quantity $\bm{\xi}\cdot\bm{u}$\footnote{For two vectors $\bm{x}$ and $\bm{y}$, we notate $\bm{x}\cdot\bm{y}=g_{\mu\nu} x^\mu y^\nu$.} varies along the congruence as \cite{Poisson:2004}
\begin{eqnarray}\label{eq:xi.u}
    \frac{D}{\ed\tau}\left(\bm{\xi}\cdot\bm{u}\right) &\equiv& \left(\bm{\xi}\cdot\bm{u}\right)_{;\nu}u^\nu\nonumber\\
    &=& \frac{1}{2}\left(\bm{u}\cdot\bm{u}\right)_{;\nu}\xi^\nu + a_{\mu;\nu}\xi^\mu u^\nu,
\end{eqnarray}
where
\begin{equation}\label{eq:a}
    a^\mu = {u^\mu}_{;\nu} u^\nu,
\end{equation}
is the four-acceleration of the non-inertial frames, according to non-gravitational effects. In this regard, a non-zero $\bm{a}$ corresponds to a vector field which is not parallel-transported along the world-lines. Accordingly, the congruence deviation equation can then be written as
\begin{eqnarray}\label{eq:Deviation_0}
    \mathfrak{A}^\mu \doteq \frac{D^2\xi^\mu}{\mathrm{d}\tau^2} &\equiv& \left({{\xi^{\mu}}_{;\nu} u^\nu}\right)_{;\gamma}u^\gamma\nonumber\\
    &=& {a^\mu}_{;\nu}\xi^\nu - 
    {R^\mu}_{\nu\alpha\beta}u^\nu\xi^\alpha u^\beta.
\end{eqnarray}
This vector, measures the \textit{relative acceleration} between two world-lines, as measured by the change in $\bm{\xi}$, and connects it to the spacetime curvature \cite{Pirani:1956,Bazanski:1989}.

According to the Eqs.~\eqref{veltau}, \eqref{mpc5} and \eqref{velphi}, we know that a congruence of charged particles with angular motion, that fall onto the charged black hole, is generated by the following four-velocity:
\begin{equation}\label{eq:umu}
    u^\mu = \left(
    \frac{E-V_q(r)}{B(r)},
    \sqrt{(E-V_-)(E-V)},
    0,
    \frac{L}{r^2}
    \right),
\end{equation}
which satisfies $\bm{u}\cdot\bm{u} = -m^2$ (we let $m=1$). The congruence deviation (Jacobi) field, related to the vector field \eqref{eq:umu}, can then take the generic form 
\begin{equation}\label{eq:xi_u}
    \xi^\mu = \left(
    \xi^0(r), \xi^1(r), 0, \xi^3(r)
    \right),
\end{equation}
for which, the consideration of the Lie transportation condition (i.e. $\mathfrak{L}_{\bm{u}}\bm{\xi}=\bm{0}$), provides
\begin{equation}\label{eq:xi-components0}
 \xi^0(r) =  2^{11/4} \lambda^2 \left(\frac{E- r V_q(r) }{4-4\lambda^2+Q^2\lambda^2}-\frac{E-V_q(r)}{r^2\left(
    4-4\lambda^2+\frac{Q^2\lambda^2}{r^2}
    \right)}\right),
\end{equation}
\begin{equation}\label{eq:xi-components1}
     \xi^1(r) = 2^{3/4} \sqrt{(E-V)(E-V_-)},
\end{equation}
\begin{equation}\label{eq:xi-components3}
     \xi^3(r) = -2^{5/4}L\left(1-\frac{1}{r^2}\right).
\end{equation}
The above vector field results in a non-zero rate of change of $\bm{\xi}\cdot\bm{u}$, indicating that the Jacobi field $\bm{\xi}$ is nowhere orthogonal to the congruence.

The four-acceleration of the infalling charged particles in electromagnetic fields, obey the following relation \cite{Misner:1973}:
\begin{equation}\label{eq:non-geodesic_1}
    a^\mu = - \frac{q}{m} g^{\mu\nu} F_{\nu\alpha} u^\alpha,
\end{equation}
which is given in terms of the field strength tensor
\begin{equation}\label{eq:F_0}
    F_{\mu\nu} = A_{\nu;\mu} - A_{\mu;\nu}.
\end{equation}
Accordingly, the congruence deviation equation \eqref{eq:Deviation_0} can be recast as (see also Refs.~\cite{Balakin2000,Heydari-Fard:2019})
\begin{equation}\label{eq:deviation_1}
    \mathfrak{A}^\mu = - {R^\mu}_{\nu\alpha\beta} u^\nu\xi^\alpha u^\beta - \frac{q}{m} g^{\mu\alpha} \left(
    F_{\alpha\beta;\nu} u^\beta \xi^\nu + F_{\alpha\beta} {u^\beta}_{;\nu} \xi^\nu
    \right).
\end{equation}
\begin{figure}[t]
    \centering
    \includegraphics[width=8cm]{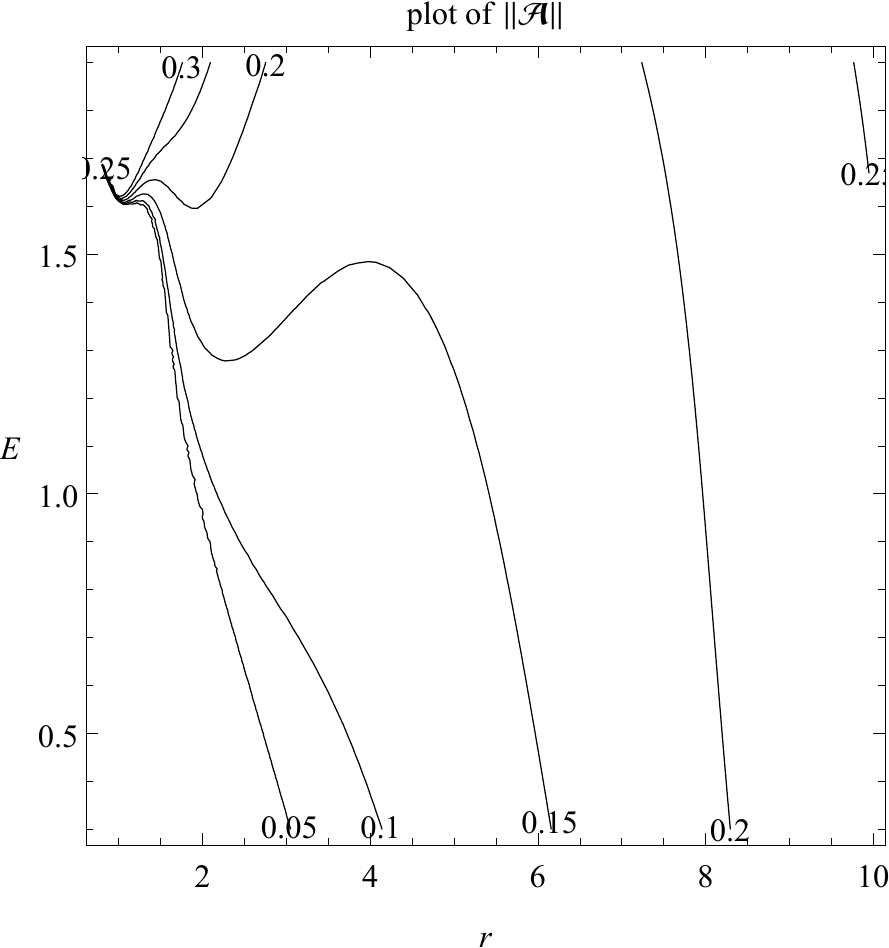}
    \includegraphics[width=8cm]{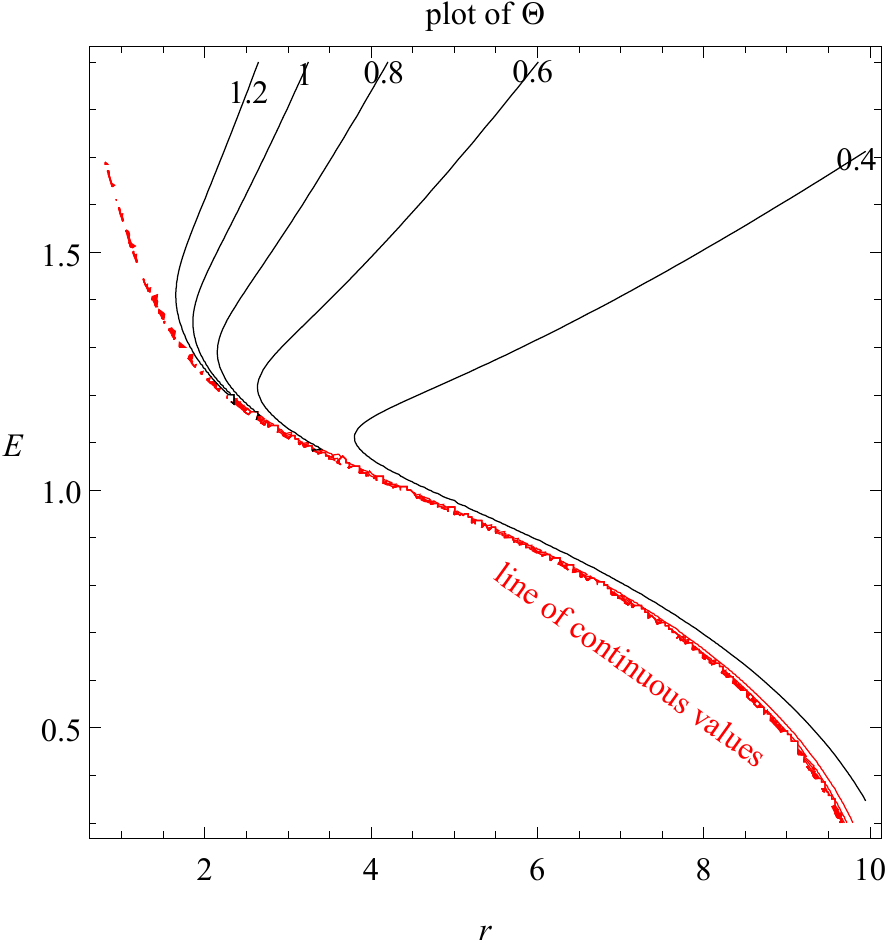}
    \caption{The behaviors of $||\bm{\mathfrak{A}}||$ and $\Theta$ for $0.3< E <1.9$, considering $Q = 1$, $q = 0.5$, $\lambda = 10$ and $L = 1$. The corresponding event and cosmological horizons are located respectively at $0.5$ and $9.98$. The contours indicate discrete values for the parameters for specific ranges of $r$ and $E$. In particular, the parameter $\Theta$, beside discrete ones, can have very close values that reside on a line tangent to the contours.}
    \label{fig:A-Theta}
\end{figure}
Since, this acceleration is related to the internal interaction of the world-lines, it naturally affects the expansion of the congruence. This expansion is defined as the fractional rate of change of the transverse subspace of the congruence, and in our case is defined as \cite{Poisson:2004}
\begin{equation}\label{eq:expansio_def}
    \Theta = {u^\mu}_{;\mu}.
\end{equation}
Accordingly, we can compare the behavior of $\bm{\mathfrak{A}}$ with that of the congruence expansion as the particle world-lines approach the black hole. For this, we consider the norm of the aforementioned vector field, i.e. $||\bm{\mathfrak{A}}||$\footnote{The norm of a vector $\bm{X}$ is defined as $||\bm{X}||\doteq\sqrt{\bm{X}\cdot\bm{X}}$.}, and plot it for a definite range of $E$, inside the causal region. Same is done for the congruence expansion (see Fig.~\ref{fig:A-Theta}). As it is seen in the figures, the approaching congruence is of positive expansion, so that its transverse cross-section increases in area and the world-lines recede from each other. This is in agreement with the positive acceleration between the world-lines, as it is shown in the diagram of $||\bm{\mathfrak{A}}||$. As the particles approach the event horizon, the congruence's internal acceleration merges to a single value at a specific $E$, which indicates that only distinct particles can reach that region and there they will maintain a constant mutual force. In other regions, distant from the event horizon, the particle deflection (and scattering) can happen under positive congruence expansion and positive internal acceleration. According to the figures, for some fixed values of $E$, the internal interactions between the world-lines remain repulsive at all distances, however, this repulsion is smaller at regions near the event horizon. This is while the congruence expansion reaches its maximum values for the same initial conditions. This is therefore a signature of scattering, where the expansion of the scattered congruence is a result of interactions with the source. On the other hand, for higher $E$, the relative acceleration and the congruence expansion take their maximum values near the black hole. The expansion in this case is naturally a result of internal interactions between the world-lines. We can therefore infer that the dynamical characteristics of a bundle of infalling world-lines on the CWBH, can indicate the effect of such interactions on the way the particles approach and recede the source, through their specific type of orbit.

%%%%%%%%%%%%%%%%%%%%% conclusion
\section{Final remarks}\label{sec:conclusion}

The gravitational effects of the electrical charge constituents of charged black holes are apparent in the motion of neutral particles in their vicinity. Charged test particles on the other hand, are also affected by an additional coulomb potential and this makes it more complicated to analyze their motion. Moving in the exterior geometry of such black holes, test particles feel this electric charge through a scalar potential, which is an external classical parameter. Despite this, the way this parameter is distributed inside the source, can change the total charge in the effective potential and therefore, can affect the particle trajectories. In general relativity, this can be inferred by applying the Einstein-Maxwell equations to obtain interior solutions of a relativistic star. In 1917, one year before proposing his theory of gravity, Weyl discovered a relationship between the metric and electrical potentials and found a class of interior axially symmetric solutions of a static source which had a quadratic dependence on the coulomb potential \cite{Weyl:1917}. This was later proved to be the case even in the absence of axial symmetry and could stabilize the source by balancing the gravitational pull and the coulomb repulsion \cite{Majumdar:1947,Papapetrou:1945}. These Weyl-type interior solutions, were then classified regarding the total integral of the interior charge density \cite{Guilfoyle:1999,Lemos:2010}, with extensions to higher dimensions \cite{Lemos:2008,Lemos:2009}. The total charge parameter of the black hole, is therefore proved to be a consequence of the type of the interior solution it obeys and this can be distinguishable regarding the particle trajectories. For example, in Ref.~\cite{Fathi:2013}, it has been shown that switching between the Weyl-type interior solutions for a RN black hole, can change the intensity and the shape of possible orbits of approaching charged test particles. 

The exterior geometry of the static charged black hole proposed by Mannheim and Kazanas in Ref.~\cite{Mannheim1991} and that in the current study, are derived from Weyl-Maxwell equations. However, besides a few Schwarzschild-like interior solutions \cite{Mannheim:1990ct,Mannheim:1991-moreSolutions}, there is not yet a study specified to the interior structure of charged relativistic stars in Weyl conformal gravity. Regarding the general interest of this paper, which was the Rutherford scattering of charged particles, such possible interior solutions could provide the chance of classifying the bending angle and the range of initial conditions to obtain a particular shape of scattering. As shown earlier in this paper, despite the similarity between the electric charge of the CWBH and that of the test particles, the scattering at some distances can be convex and attractive, which indicates the unbalance between the gravitational and coulomb potentials felt by the test particles. If the source is endowed with a particular charge density function and definite interior solutions, then the scattering can be categorized in accordance with the total charge integral for each of the solutions. So, an outlook for future studies can be looking for obtaining charged interior solutions to Weyl conformal gravity and matching them with the exterior geometry of the respected charged black holes. This way, beside categorizing the types of motion of test particles, we will also be able to investigate them around a star under gravitational collapse.

%%%%%%%%%%%%%%%%%%
\begin{acknowledgements}
	M. Fathi has been supported by the Agencia Nacional de Investigaci\'{o}n y Desarrollo (ANID) through DOCTORADO Grant No. 2019-21190382, and No. 2021-242210002. J.R. Villanueva was partially supported by the Centro de Astrof\'isica de Valpara\'iso (CAV).
\end{acknowledgements}

%%%%%%%%%%%%%%%%%%%%%%%%%%%%%%%%%%%%%%%%%%%%%%%%%%%
\appendix

%%%%%%%%%%%%%%%%%
\section{The method of finding $L_U$ in Eq.~\eqref{T.25}}\label{app:A0}

Equation.~\eqref{eq:Vprime=0} allows for obtaining an expression for $L_U$, by solving
\begin{equation}\label{C.23} \mathfrak{a}L_{U}^{4}-\mathfrak{b} L_{U}^{2}+\mathfrak{c}=0,
\end{equation}
in which
\begin{subequations}
\begin{align}
&\mathfrak{a}={(Q^2-2r_U^2)^2\over r_U^6 },
\\
&\mathfrak{b}={2Q^2(1+q^2)\over r_U^2}-{Q^4(2+q^2)\over 2r_U^4}-{8r_U^2-2Q^2(1-q^2)\over \lambda^4},
\\
&\mathfrak{c} ={Q^4(1+2q^2)\over 4r_U^2}-2q^2Q^2+{4r_U^6\over \lambda^4}-{2Q^2r_U^2(1-q^2) \over \lambda^2}.
\end{align}
\end{subequations}
Solving Eq.~\eqref{C.23} for $L_{U}^2$, then yields the value in Eq.~\eqref{T.25}.

%%%%%%%%%%%%%%%%%%%%%%%%%
\section{Finding the angular equation of motion}\label{app:angular}

Since the closest approach happens at $r_S$, to deal with the integral in Eq.~\eqref{phi1}, we define the following non-linear change of variable:
\begin{equation}\label{cv1}
	u \doteq {1\over  \frac{r}{r_S}-1},
\end{equation}
which reduces Eq.~\eqref{phi1} to
 \begin{equation}\label{phi1b}
\phi(r) =\kappa_0\left[\int_{u}^{\infty}{\ed u\over  \sqrt{\mathcal{P}_3(u)} }-u_3\int_{u}^{\infty}{\ed u\over  (u+u_3)\sqrt{\mathcal{P}_3(u)} }\right],
\end{equation}
where $u_j\doteq {1\over (r_j/r_S)-1}$, with $j=\{2,3,5,6\}$, and
\begin{equation}\label{eqa}
\mathcal{P}_3(u)\equiv u^3+\mathbf{a} u^2+\mathbf{b} u+\mathbf{c},
\end{equation}
with
\begin{subequations}
\begin{align}\label{abc}
&\mathbf{a} = u_2+u_5+u_6,\\
&\mathbf{b} = u_2(u_5+u_6)+u_5 u_6,\\ 
&\mathbf{c} = u_2 u_5 u_6.
\end{align}
\end{subequations}
Defining
\begin{equation}
    \kappa_0= {\upsilon \over r_S^2} u_3 \sqrt{u_2 u_5 u_6},
\end{equation}
and applying another change of variable 
\begin{equation}\label{eq:c-u}
U\doteq\frac{1}{4}\left(
u+\frac{\mathbf{a}}{3}
\right),
\end{equation}
we can rewrite Eq.~\eqref{phi1b} as
 \begin{equation}\label{phi1c0}
\phi(r) =\kappa_0\left[\int_{U}^{\infty}{\ed U\over  \sqrt{\mathit{P}_3(U)} }-{u_3\over 4}\int_{U}^{\infty}{\ed U\over  (U+U_3)\sqrt{\mathit{P}_3(U)} }\right],
\end{equation}
given that $U_3 = \frac{1}{4}\left(u_3+\frac{\mathbf{a}}{3}\right)$, and
\begin{equation}\label{eqa}
\mathit{P}_3(u)\equiv 4 U^3-\mathbf{g}_2 U-
\mathbf{g}_3.
\end{equation}
Direct integration of Eq.~\eqref{phi1c0}, results in the expression in Eq.~\eqref{phi1c}.

%%%%%%%%%%%App A
\section{Solving depressed quartic equations}\label{app:Ap}

The condition $V_r'(r)=0$, provides the following equation of eighth degree:
\begin{equation}\label{eq:unstable_radial_0}
r^8+\tilde{a} r^4+\tilde{b} r^2+\tilde{c}=0.
\end{equation}
To solve this equation, we firstly make the change of variable $r^2\doteq x$. Afterwards, we combine the methods of Ferrari and Cardano to solve a  depressed quartic equation of the form (originally studied by Cardano in Ref.~\cite{Cardano:1993})
\begin{equation}\label{a1}
x^4+\tilde{a} x^2+\tilde{b} x+\tilde{c}=0,~~~~(\tilde{a},\tilde{b},\tilde{c}) \in \mathbb{R}. 
\end{equation}
This equation can be rewritten as the product of two quadratic equations, as follows:
\begin{equation}\label{a2}
x^4+\tilde{a} x^2+\tilde{b} x+\tilde{c} = (x^2-2 \tilde\alpha x+\tilde\beta)(x^2+2  \tilde\alpha x+\tilde\gamma)=0.
\end{equation}
Accordingly, we obtain
\begin{subequations}
\begin{align}
&\tilde{a}=\tilde\beta+\tilde\gamma-4\tilde\alpha^2,  \\
&\tilde{b}=2\tilde\alpha(\tilde\beta-\tilde\gamma),\\
&\tilde{c}=\tilde\beta \tilde\gamma\label{eq:app_cbeta}.
\end{align}
\end{subequations}
Solving the first two equations for $\tilde\beta$ and $\tilde\gamma$, yields
\begin{subequations}
\begin{align}
&\tilde\beta=2\tilde\alpha^2 +{\tilde{a}\over 2}+{\tilde{b}\over 4\tilde\alpha},\\
&\tilde\gamma=2\tilde\alpha^2 +{\tilde{a}\over 2}-{\tilde{b}\over 4\tilde\alpha},
\end{align}
\end{subequations}
which together with Eq.~\eqref{eq:app_cbeta}, results in an equation of sixth degree in $\tilde\alpha$:
\begin{equation}
\tilde\alpha^6+{\tilde{a}\over 2}\tilde\alpha^4+\left( {\tilde{a}^2\over 16}-{\tilde{c}\over 4}\right) \tilde\alpha^2-{\tilde b^2\over 64}=0.
\end{equation}
Applying the change of variable
\begin{equation}
    \tilde\alpha^2=\tilde{U}-{\tilde{a}\over 6},
\end{equation}
we obtain the depressed cubic equation
\begin{equation}
\tilde{U}^3-\tilde\eta_2 \,\tilde{U}-\tilde\eta_3=0,
\end{equation}
where
\begin{subequations}
\begin{align}
&\tilde\eta_2={\tilde{a}^2\over 48}+{\tilde{c}\over 4},\\
&\tilde\eta_3={\tilde{a}^3\over 864}+{\tilde{b}^2\over 64}-{\tilde{a} \tilde{c}\over 24}.
\end{align}
\end{subequations}
The real solution to this cubic equation is obtained as \cite{Nickalls:2006,Zucker:2008} 
\begin{equation}
\tilde{U}= 2\sqrt{{\tilde\eta_2\over 3}} \cosh\left(\frac{1}{3}\arccosh\left({3\over 2}\tilde\eta_3\sqrt{{3\over \tilde\eta_2^3}}  \right) \right).
\end{equation}
Therefore, the roots of Eq.~\eqref{a1} are 
\begin{subequations}
\begin{align}
& x_1=\tilde\alpha+\sqrt{\tilde\alpha^2-\tilde\beta},\\
& x_2=\tilde\alpha-\sqrt{\tilde\alpha^2-\tilde\beta},\\
&  x_3=-\tilde\alpha+\sqrt{\tilde\alpha^2-\tilde\gamma},\\
&  x_4=-\tilde\alpha-\sqrt{\tilde\alpha^2-\tilde\gamma}.
\end{align}
\end{subequations}

%%%%%%%%%%%%%%%%%%%%%%%%%%%%%
\section{Solving the equation of motion for frontal scattering}\label{app:A1}

Equation \eqref{vel1} can be recast as
\begin{equation}\label{eq:drdtau-rad}
\left(\frac{{\rm d}r}{{\rm d}\tau}\right)^{2}=\frac{m^2 \left(r-r_s\right) \mathfrak{p}_3(r)}{\lambda^2 r^2},
\end{equation}
in which 
\begin{equation}\label{eq:mathfrakp3}
 \mathfrak{p}_3(r)\equiv r^3+r_s r^2+(r_s^2+\bar{a})r+r_s^3+r_s\bar{a}+\bar{b}.
\end{equation}
Considering $r_s$ as the initial position, we can rewrite Eq.~\eqref{eq:drdtau-rad} as
\begin{equation}\label{eq:tau(r)-int-0}
\tau (r)={\lambda \over m}\int_{r_s}^{r} {r \ed r \over \sqrt{(r-r_s) \mathfrak{p}_3(r)}},
\end{equation}
which by the linear change of variable
\begin{equation}\label{eq:c-x}
	z \doteq {r \over r_s}-1,
\end{equation}
reduces to
\begin{equation}\label{eq:tau(r)-int-1}
\tau (z)={\lambda \over m}\int_{0}^{z} {(z+1) \ed z \over \sqrt{z \tilde{\mathfrak{p}}_3(z)}},
\end{equation}
where
\begin{equation}\label{eq:mathfrakp3tilde}
    \tilde{\mathfrak{p}}_3(z)\equiv z^3 +4 z^2 +\gamma_1 z+\gamma_0,
\end{equation}
and
\begin{subequations}\label{gamma12}
\begin{align}
 &\gamma_1=6+\frac{\bar{a}}{r_s^2},\\
 &\gamma_0=4+\frac{2\bar{a}}{r_s^2}+\frac{\bar{b}}{r_s^3}.
\end{align}
\end{subequations}
Now, letting
\begin{equation}\label{uz}
	u\doteq\frac{1}{z},
\end{equation}
yields the following reduced integral form of Eq.~\eqref{eq:tau(r)-int-1}:
\begin{equation}\label{eq:tau(u)-int-2}
\tau (u)={-\lambda \over m\sqrt{\gamma_0}}\left( 
\int_{\infty}^{u} {\ed u\over \sqrt{\bar{\mathfrak{p}}_3(u)}}+
\int_{\infty}^{u} {\ed u \over u \sqrt{\bar{\mathfrak{p}}_3(u)}}
\right), 
\end{equation}
in which
\begin{equation}\label{eq:mathfrakp3-tilde-u}
     \bar{\mathfrak{p}}_3(u)\equiv u^3 +\frac{\gamma_1}{\gamma_0} u^2 +\frac{4}{\gamma_0} u+\frac{1}{\gamma_0}.
\end{equation}
Applying the last change of variable
\begin{equation}\label{eq:c-u}
	u\doteq 4 \mathrm{U}-\frac{\gamma_1}{3\gamma_0},
\end{equation}
we get
\begin{equation}\label{eq:tau(U)-int-3}
	\tau (\mathrm{U})={-\lambda \over m\sqrt{\gamma_0}}\left( 
	\int_{\infty}^{\mathrm{U}} {\ed\mathrm{U}\over \sqrt{\bar{\mathfrak{P}}_3(\mathrm{U})}}\right.\\
	\left.+{1\over 4}	\int_{\infty}^{\mathrm{U}} {\ed\mathrm{U} \over \rm (U- {\gamma_1\over 12  \gamma_0}) \sqrt{\bar{\mathfrak{P}}_3(\mathrm{U})}}
	\right), 
\end{equation}
where we have defined
\begin{equation}\label{eq:mathfrakp3-tilde-U}
    \bar{\mathfrak{P}}_3(\mathrm{U})\equiv 4\mathrm{U}^3 - \bar{g}_2 \mathrm{U}^2 -\bar{g}_3.
\end{equation}	
The direct integration of the elliptic integral in Eq.~\eqref{eq:tau(U)-int-3}, now results in the expression in Eq.~\eqref{eq:tau(r)-int-4}.

%%%%%%%%%%%%%%%%%%%%%%%%%%%%%%%%%%%%%%%%%%%%

%\bibliographystyle{spphys} 
\bibliography{Biblio_v1.bib}
%\end{thebibliography}

\end{document}